\documentclass[12pt]{iopart}

\usepackage[T1]{fontenc}
\usepackage{color}
 \expandafter\let\csname equation*\endcsname\relax
 \expandafter\let\csname endequation*\endcsname\relax
\usepackage{amsmath}
\usepackage{bbm}

\input epsf
\usepackage{iopams}  

\usepackage[pdftex]{graphicx}

\newcommand{\bref}[1]{(\ref{#1})}

\newcommand{\bS}{{\bf \hat S}}
\newcommand{\itext}[1]{\text{\it #1}}

\begin{document}

\title[Antiferromagnetism, charge density wave, and $d$-wave superconductivity...]{Antiferromagnetism, charge density wave, and $d$-wave superconductivity in the extended $t$--$J$--$U$ model: role of intersite Coulomb interaction and a critical overview of renormalized mean field theory}

\author{M Abram$^1$, M Zegrodnik$^2$, and J Spa\l ek$^{1*}$}

\address{$^1$Marian Smoluchowski Institute of Physics, Jagiellonian University, {\L}ojasiewicza  11, 30-348 Krak\'ow, Poland\\
$^2$AGH University of Science and Technology,
Faculty of Physics and Applied Computer Science, Al. Mickiewicza 30,
30-059 Krak\'{o}w, Poland}

\ead{*jozef.spalek@uj.edu.pl}
\begin{abstract}
In the first part of the paper, we study the stability of antiferromagnetic (AF), charge density wave (CDW), and superconducting (SC)
states within the $t$-$J$-$U$-$V$ model of strongly correlated electrons by
using the \textit{statistically consistent Gutzwiller approximation} (SGA). 
We concentrate on the role of the intersite Coulomb interaction term $V$ in stabilizing the CDW phase.
In particular, we show that the charge ordering appears only above a critical value of $V$ in a limited hole-doping range $\delta$.
The effect of the $V$ term on SC and AF phases is that a strong interaction suppresses SC,
whereas the AF order is not significantly influenced by its presence.
In the second part, separate calculations for the case of pure SC phase have been carried out within an extended approach (the \textit{diagrammatic expansion for the Gutzwiller wave function}, DE-GWF)
in order to analyze the influence of the intersite Coulomb repulsion on the SC phase with the higher-order corrections included beyond the SGA method.
In the Appendices we discuss the ambiguity connected with the choice of the Gutzwiller renormalization factors within the renormalized mean filed theory when either AF or CDW orders are considered. At the end we overview briefly the possible extensions of the current models to make description of the SC, AF, and CDW states on equal footing. 
\end{abstract}


\submitto{\JPCM (2017)\\ \emph{Published:} \JPCM {\bf 29}, 365602 (2017), DOI:10.1088/1361-648X/aa7a21}
\maketitle

\section{Introduction}\label{sec:intro}
It is generally accepted that the copper-oxide planes contain strongly correlated electronic states and those are instrumental for achieving the high-temperature superconductivity. The simplest models describing those states are the single-band $t$-$J$, Hubbard, and related extended models \cite{Anderson1988,Zhang1988-PhysRevB.37.3759,Anderson2004,Ogata2008-RepProgPhys,Randeria2012} . A standard treatment in this situation is the renormalized mean-field theory (RMFT) which provides description of principal superconducting properties in a qualitative manner \cite{Zhang1988_2,Edegger2007}. In addition, variational Monte-Carlo (VMC) approach provides a semiquantitative description of selected properties, even through the analysis limited to the systems of finite size \cite{Randeria2012,Edegger2007,Randeria2001}. Very recently, we have carried out \cite{Spalek-arxiv} an extensive analysis of the universal properties of the cuprates for the case pure superconducting solution. As a result we have shown that a very 
good agreement with principal experimental data can be achieved when including the higher-order terms within the diagrammatic expansion of the Gutzwiller wave function method (DE-GWF) for the case of the $t$-$J$-$U$ model. In such a picture the intraatomic (Hubbard) interaction magnitude $U$ is not regarded as extremely strong
which means that the limit $U/t\rightarrow \infty$ is not assumed at the start. In this situation,
a straightforward decomposition of the narrow-band states into the Hubbard subbands, with the upper subband
(for the band filling $n \leqslant 1$)
being unoccupied, is not physically crucial and in effect, the $t$-$J$ model does not follow directly from the perturbation
expansion of the Hubbard model in powers of $t$/$U$ \cite{Chao1977-JournPhysC,Chao1978-PhysRevB.18.3453,Spalek1977-PhysicsBC,Spalek1978-PhysStatusSolidiB,Spalek1981-PhysStatusSolidiB,Spalek2007-ActaPhisPol}. Instead, the antiferromagnetic kinetic exchange interaction arises
from the superexchange via 2$p_{\sigma}$ states due to oxygen \cite{Jefferson1992,Plakida2015-Book}. Consequently, small nonzero number of double occupancies is admissible for the non-half-filled band case and in such a situation the Hubbard term $\sim U$ 
provides a direct contribution to the states.
This argument justifies the generalization of the $t$--$J$ model
to the $t$--$J$--$U$ form. Originally, the Hubbard term has been added to the $t$--$J$ model when introducing
the so-called Gossamer superconductivity (for detailed discussion cf. Ref.~\cite{Abram2013-PhysRevB.88.094502} and papers cited therein;
cf. also Ref.~\cite{Zhang2003-prl}).

Motivated by our previous results for the case of pure SC phase \cite{Spalek-arxiv} and the antiferromagnetic phase \cite{Abram2013-PhysRevB.88.094502}, we focus here on the one-band description of the Cu-O plane within the $t$-$J$-$U$-$V$ approach and analyze the three most important phases related to the copper-based high temperature superconducting (HTS) compounds:
antiferromagnetic (AF), interunitcell charge ordered (CDW), and superconducting (SC) together. We mainly concentrate on the formation of the CDW state induced by the intesite Coulomb repulsion $\sim V$ as the most natural factor determining its appearance \cite{Eder1996-PhysRevB.54.R732}. The inclusion of the $V$-term leads to the generalized $t$-$J$-$U$-$V$ model. We also analyze the influence of the modulation vector $\mathbf{Q}$ on the charge ordering stability region in the phase diagram.

The interest in the field of charge ordered (CO) states within the one-band models of Cu-O plane has been revived in the recent years due to growing experimental evidence that the charge ordering appears spontaneously or in the presence of magnetic field in the underdoped region of high temperature cuprate 
superconductors \cite{Keimer2015-Nature,Wu2011-Nature,Wu2015-Nature,Comin2014-Science,Silva2015-Science,Lubashevsky2014-PRL,Achkar2012-PRL,Tranquada1995-Nature,Attfield2006-Solid}. For the La-based materials the NMR experiments suggested a commensurate stripe like charge order with a period of 4 unit cells \cite{Wu2011-Nature}. For YBCO and Bi-based materials it has been found that the CDW phase is characterized by the modulation vectors $(0,Q)$ and $(Q,0)$ \cite{Wu2015-Nature,Achkar2012-PRL,Ghiringhelli2012-Science,Chang2012-Nature} which are incommensurate with a weakly doping dependent period of $\sim3.1$ for YBCO \cite{Achkar2012-PRL,Ghiringhelli2012-Science,Blackburn2013-PRL} and $Q\sim 2.6$  ($0.3$) for the single-layer (double-layered) Bi-based compounds \cite{Comin2014-Science,daSilva2014-Science}. Another important conclusion is that, regardless of the details of the charge ordered phase, a competition between superconductivity (SC) and CDW takes place in the cuprates. Moreover, $T_{CDW}$ appears to 
be 
larger than $T_C$. Some analyses indicate that the charge order is located predominantly on the bonds connecting the Cu sites \cite{
Achkar2013-PRL,Sachdev2003-RMP}. Such a scenario has been considered theoretically by Allais et al \cite{Allais-Phys.Rev.B-90.155114} by assuming a proper $d$-form factor of the CDW which leads to the modulation vector close to that observed experimentally \cite{Comin2014-Science,Silva2015-Science}. The $d$-wave symmetry of the CDW ordering has also been considered in Refs. \cite{Fujita2014-PNAS, Fradkin2015-RevModPhys.87.457} to preserve the nodal nature of the $k_x=k_y$ direction at low temperature. 

The research in this field is additionally motivated by the fact that there might be a connection between the pseudogap in the cuprates and the charge ordering \cite{Chang2012-Nature, Wise2008-NatPhys.4.696, Shen-Science}. It has also been suggested that
for certain materials the CDW state may have a three-dimensional character in nonzero applied magnetic field \cite{Gerber2015-Science,Julien2015-Science}. Furthermore, the appearance of the so-called
pair-density-wave state has been proposed which can coexist with the CDW state
and may lead to the appearance of the pseudogap anomaly \cite{Fradkin2015-NaturePhys,Lee-Phys.Rev.X-4.031017,Wang-Phys.Rev.Lett-114.197001}. In such a state the SC order parameter has a nonzero Cooper pair momentum similar to that appearing in the Fulde-Ferrell-Larkin-Ovchinnikov phase.

What concerns the theoretical analysis, various calculation schemes have been applied to the (extended) Hubbard, $t$-$J$, and related models in seeking the stability of CO states \cite{Allais-Phys.Rev.B-90.155114, Amaricci-Phys.Rev.B-82.155102,Sachdev-Phys.Rev.Lett-111.027202,Corboz-Phys.Rev.Lett-113.046402, Caprara-arxiv, Kapcia2015-JSNM,Kapcia2016-arxiv}. In contrast to some of those considerations, it has been suggested very 
recently \cite{Tu2016-ScientificReport} that the CDW solution for the $t$--$J$ model has always a slightly higher energy
than the generic SC+AF solution. It is also not settled if the simplified one-band approach can lead to the proper description of the charge-ordered phase in the cuprates. The STM experiments \cite{Mesaros2011-Science} point to the inraunitcell charge order which would require a 3-band model of the Cu-O plane for a realistic theoretical analysis.

In the first part of the paper, our analysis is carried out with
the use of the so-called \emph{statistically consistent Gutzwiller approximation} (SGA),
within which we can account for electron correlations in a reasonable computing time
(for the derivation of SGA method see Refs.~\cite{SGA,Jedrak2010} and for its applications see Refs.~\cite{Abram2013-PhysRevB.88.094502,Kaczmarczyk2011,Zegrodnik2013-JPCM,Kadzielawa2013,Wysokinski2014prb, Abram2016}). This approximation represents a consistent version of the renormalized mean field theory (RMFT), as explained in the papers just mentioned.
In the second part of the paper (cf. Sec. \ref{sec:4.3}),
we test the influence of the intersite Coulomb interaction on the robustness of the pure superconducting solution beyond SGA with the use of a \emph{systematic diagrammatic expansion of the Gutzwiller-wave function} (DE-GWF method) (for details of this approach see Ref. \cite{Bunemann2012} and for the application of this method to the $t$-$J$-$U$ model see \cite{Spalek-arxiv,Zegrodnik2017-PRB}).
This last approach allows us to go beyond the SGA in a systematic manner
and to take into account non-local correlations in higher orders of the expansion.
The differences between the SGA and the DE-GWF solutions are specified there.
Additionally, in \ref{app:GutzwillerFactors}, we discuss an inherent ambiguity in choosing the Gutzwiller
renormalization factors when either AF or CDW states are considered within SGA.
Namely, we show that different calculation schemes used in the literature lead to 
different forms of the Gutzwiller factors, what results in
different stability regimes of the AF phase.

\section{$t$--$J$--$U$--$V$ model}

The starting Hamiltonian for the subsequent analysis has the following form,
\begin{multline}\allowdisplaybreaks
\mathcal{\hat H}_{\text{$t$--$J$--$U$--$V$}} =
t \;\sum_{\langle i,\, j \rangle,\, \sigma} \left( \hat c^\dagger_{i\sigma} \hat c_{j\sigma} + \mathnormal{H.c.} \right)
 + t' \; \sum_{\langle\langle i,\, j \rangle\rangle,\, \sigma}  \left( \hat c^\dagger_{i\sigma} \hat c_{j\sigma} + \mathnormal{H.c.} \right)\\
 + J \sum_{\langle i,\, j \rangle} \hat {\bf S}_i \cdot \hat {\bf S}_j
 + U \sum_{i} \hat n_{i\uparrow} \hat n_{i\downarrow}
 + \underbrace{\left( \tilde V - \frac{1}{4} J \right)}_{V} \sum_{\langle i,\, j \rangle, \sigma, \sigma'}\!\!\!\! \hat n_{i\sigma} \hat n_{j\sigma'},
 \label{eq:t-J-V_def1} 
\end{multline}
% \info{Do tej pory sumę $\sum_{\langle i,\, j \rangle,\, \sigma}$ zapisywaliśmy jako sumę po \emph{unikalnych}
% parach, tj. jeśli sumowaliśmy po $(i,j)$, to \emph{nie} sumowaliśmy po $(j,i)$, por. J.~Jędrak and J.~Spałek PRB {\bf 81}, 073108 (2010)
% oraz późniejsze prace, w tym \cite{Abram2013-PhysRevB.88.094502}.
% W takim ujęciu, powyższy wzór jest poprawny. Jeśli jednak chcielibyśmy to tutaj zmienić, tj. $\sum_{\langle i,\, j \rangle}$
% miałoby być tu sumą po wszystkich parach z powtórzeniami (każda para liczona podwójnie), to wtedy powyższy wzór
% powinien być zmieniony na:
% \begin{multline*}\allowdisplaybreaks
% \mathcal{\hat H}_{\text{$t$--$J$--$U$--$V$}} =
% t \!\!\!\! \sum_{\langle i,\, j \rangle,\, \sigma} \!\!\!  \hat c^\dagger_{i\sigma} \hat c_{j\sigma} 
%  + t' \!\!\!\!\!\! \sum_{\langle\langle i,\, j \rangle\rangle,\, \sigma} \!\!\! \hat c^\dagger_{i\sigma} \hat c_{j\sigma}
%  + \frac{J}{2} \sum_{\langle i,\, j \rangle} \hat {\bf S}_i \cdot \hat {\bf S}_j \\
%  + U \sum_{i} \hat n_{i\uparrow} \hat n_{i\downarrow}
%  + \frac{1}{2} \underbrace{\left( \tilde V - \frac{1}{4} J \right)}_{V} \sum_{\langle i,\, j \rangle, \sigma, \sigma'}\!\!\!\! \hat n_{i\sigma} \hat n_{j\sigma'},
%  \label{eq:t-J-V_def1} 
% \end{multline*}}
where $\sum_{\langle i,\, j \rangle}$ and $\sum_{\langle\langle i,\, j \rangle\rangle}$ denote summation
up to all nearest and second nearest neighbors. Furthermore,
$t$ and $t'$ are respectively the hopping amplitudes between the nearest and the next nearest neighboring sites,
$J$ is the antiferromagnetic exchange integral, and $U$ ($\tilde V$) is the onsite (intersite) Coulomb repulsion magnitude.
The standard notation is used, where $\hat c_{i\sigma}^\dagger$ and $\hat c_{i\sigma}$
are, respectively, the creation and the annihilation operators, for electron with spin quantum number $\sigma= \pm 1$ located at site~$i$.
Similarly, $\hat n_{i\sigma} \equiv \hat c_{i\sigma}^\dagger \hat c_{i\sigma}$ and $\bS_i \equiv (\hat S_{i}^+,\, \hat S_{i}^-,\, \hat S_{i}^z)$,
where $\hat S_{i}^\sigma \equiv \hat c_{i\sigma}^\dagger \hat c_{i\bar\sigma}$, and
$\hat S_{i}^z \equiv \frac{1}{2} \left(  \hat n_{i\uparrow} - \hat n_{i\downarrow} \right)$.

As has already been mentioned, the appearance of the $J$ term in this approach is attributed mainly to the $d$--$d$ superexchange
via 2$p_{\sigma}$ states due to oxygen \cite{Jefferson1992,Plakida2015-Book}. The finite value of the Coulomb repulsion $U$
leads to a relatively small but nonzero population of the upper Hubbard subband \cite{Zhang1988-PhysRevB.37.3759, Zhang2003-prl, Laughlin2006-PhilMag}.
In such a situation the appearance of both the $J$ and the $U$ terms is physically admissible in the Hamiltonian. 
For $V=0$ and when $U\rightarrow \infty$, the limit of the $t$--$J$ model is recovered.
On the other hand, for $J=V=0$, we obtain the limit of the Hubbard model.
Nevertheless, the model is not only constructed as a formal generalization of those two limits. As can be seen from the numerous estimates
of the model parameters, the typical values of the parameters of the one-band model are: $t=-0.35$~eV, $t'=0.25|t|$ and $U\approx8$--$10$~eV,
so that the ratio of $U$ to the bare bandwidth $W = 8|t|$ is $U/W \approx 2.5$--$3$, i.e., only by the factor of about two higher
than typical required for Mott-Hubbard localization in the limit of half-filled band\cite{Gebhard1997-Book}.
As a consequence, the Hubbard gap is $U-W \sim W$ and the double occupancy can be estimated as $d^2 \lesssim \frac{t}{U} \delta \sim 10^{-2}$,
where $\delta$ is the hole doping. Additionally, the value of $U$ is reduced to the value $U-V \sim \frac{2}{3} U$
when the $V$ term is present \cite{Chao1977-JournPhysC}.
Also, as said above, the value of $J$ cannot be regarded as resulting from the $t$/$U$ expansion of the (extended)
Hubbard model \cite{Chao1977-JournPhysC, Chao1978-PhysRevB.18.3453, Spalek1977-PhysicsBC, Spalek1978-PhysStatusSolidiB, Spalek1981-PhysStatusSolidiB, Spalek2007-ActaPhisPol, NovaScience-SpalekHonig}, and both $J$ and $U$ can be regarded as practically independent variables.
The last term comes partially from the derivation of $t$--$J$ model from the Hubbard model
(cf. \cite{Chao1977-JournPhysC, Chao1978-PhysRevB.18.3453, Spalek1977-PhysicsBC, Spalek1981-PhysStatusSolidiB, Spalek2007-ActaPhisPol, NovaScience-SpalekHonig,Spalek1988-PhysRevB.37.533}) and partially (the part $\sim \tilde V$) represents an explicit
intersite Coulomb repulsion of electrons located on the nearest neighboring sites.
For simplicity, we denote $ V \equiv \tilde V - \frac{1}{4}J$.

\section{Methods}

\subsection{Statistically Consistent Gutzwiller Approximation (SGA)}
In this subsection we describe the principles of the statistically consistent Gutzwiller approximation (SGA) \cite{SGA, Jedrak2010, Gutzwiller1962, Gutzwiller1965,Jedrak2011} which we use to solve the Hamiltonian \bref{eq:t-J-V_def1}. The phase which are of interest in our analysis are: the paired phase, the antiferromagnetic phase, and the charge ordered phase. We assume that the two last phases are in the simplest two-sublattice form with the modulation vector $\mathbf{Q}=(\pi, \pi)$. However, at the end we also show some results for the case of CDW phase with $\mathbf{Q}=(2\pi /3, 0)$. 

The main idea behind the Gutzwiller approach is to express the wave function of the system in the following manner
\begin{equation}
|\Psi\rangle = \hat P\, |\Psi_0\rangle \equiv \prod_{i}^N \hat P_i\, |\Psi_0\rangle,
\label{eq:Gutzwiller_wf}
\end{equation}
where the correlator $\hat P_i$ in its most general form is as follows:
\begin{equation}
\hat P_i = \sum_{\Gamma} \lambda_{i,\Gamma} | \Gamma\rangle_{i\,i} \langle \Gamma|.
 \label{eq:P-general_1}
\end{equation}
The variational parameters $\lambda_{i,\Gamma}\in\{\lambda_{i\emptyset},\lambda_{i\uparrow},\lambda_{i\downarrow},\lambda_{i\uparrow\downarrow}\}$
weight the configurations corresponding to states from the local basis: $|\emptyset\rangle_i$, $|\!\!\uparrow\rangle_i$, $|\!\!\downarrow\rangle_i$, and $|\!\!\uparrow\downarrow\rangle_i$, respectively.
The non-correlated wave function, $|\Psi_0\rangle$, is taken as the broken-symmetry state of our choice.
 It has been shown by B\"unemann et al.\ \cite{Bunemann2012}, that it is convenient to choose the $\hat{P}_i$ operator that fulfills the following relation,
\begin{equation}
 \hat{P}_i^2=1+x_i\hat{d}^{\,\textrm{HF}}_i,
 \label{eq:P2_diag}
\end{equation}
where $x_i$ is yet another variational parameter and $\hat{d}^{\,\textrm{HF}}_i=\hat{n}_{i\uparrow}^{\textrm{HF}}\hat{n}_{i\downarrow}^{\textrm{HF}}$,
where $\hat{n}_{i\sigma}^{\textrm{HF}}=\hat{n}_{i\sigma}-n_{i\sigma}$ with $n_{i\sigma}=\langle\Psi_0|\hat{n}_{i\sigma}|\Psi_0\rangle$. The $x_i$ parameter can be directly connected with the double occupancy probability $d_i^2=\langle\hat{n}_{i\uparrow}\hat{n}_{i\downarrow}\rangle$
\begin{equation}
 x_i \equiv \frac{d_i^2 - n_{i\uparrow} n_{i\downarrow}}{n_{i\uparrow} n_{i\downarrow} (1 - n_{i\uparrow}) (1 - n_{i\downarrow})}.
\end{equation}
The $\lambda_{i,\Gamma}$ parameters can be expressed with the use of $x_i$ or $d_i$, as we show in Appendix A, therefore we are left always with only one local variational parameter.

As the operator $\hat{P}$ is in general not unitary, the expectation value of the ground-state energy of the system is expressed as follows,
\begin{equation}
E \equiv\langle\mathcal{\hat H}\rangle\equiv \frac{\langle\Psi|\mathcal{\hat H}|\Psi\rangle}{\langle\Psi|\Psi\rangle} =
\frac{\langle\Psi_0|\hat P \mathcal{\hat H} \hat P|\Psi_0\rangle}{\langle\Psi_0| \hat P^2 |\Psi_0\rangle}
\approx \langle \Psi_0 | \mathcal{\hat H}_{\itext{eff}} | \Psi_0 \rangle.
%\equiv \langle\mathcal{\hat H}_{\itext{eff}}\rangle_0,
\label{eq:rownowaznoscRMFT_GA}
\end{equation}
In other words, instead of calculating the average of the initial Hamiltonian \bref{eq:t-J-V_def1} with respect to
usually complicated, many-particle, wave function $|\Psi \rangle$, we choose to modify that Hamiltonian
(presumably by making it more complicated)
in order to have the relatively simple task of calculating its average with respect to the wave function $|\Psi_0 \rangle$
represented by a single Slater determinant. Within the SGA approach we make use of the following approximations while calculating $E$
\begin{equation}
 \langle\Psi|\hat{o}_{i} \hat{o}^{\prime}_{j}|\Psi \rangle\approx \langle\Psi_0|\hat{P}_i\hat{o}_{i}\hat{P}_i\hat{P}_j \hat{o}^{\prime}_{j}\hat{P}_j|\Psi_0 \rangle,\quad \langle\Psi|\hat{o}_{i}|\Psi \rangle\approx\langle\Psi_0|\hat{P}_i\hat{o}_{i}\hat{P}_i|\Psi_0 \rangle,
 \label{eq:GA}
\end{equation}
for any two local operators $\hat{o}_{i}$ and $\hat{o}^{\prime}_{j}$ from our Hamiltonian. Such relations are exact in the infinite dimensions limit.

\begin{figure}
 \centering
 \includegraphics[width=0.8\textwidth]{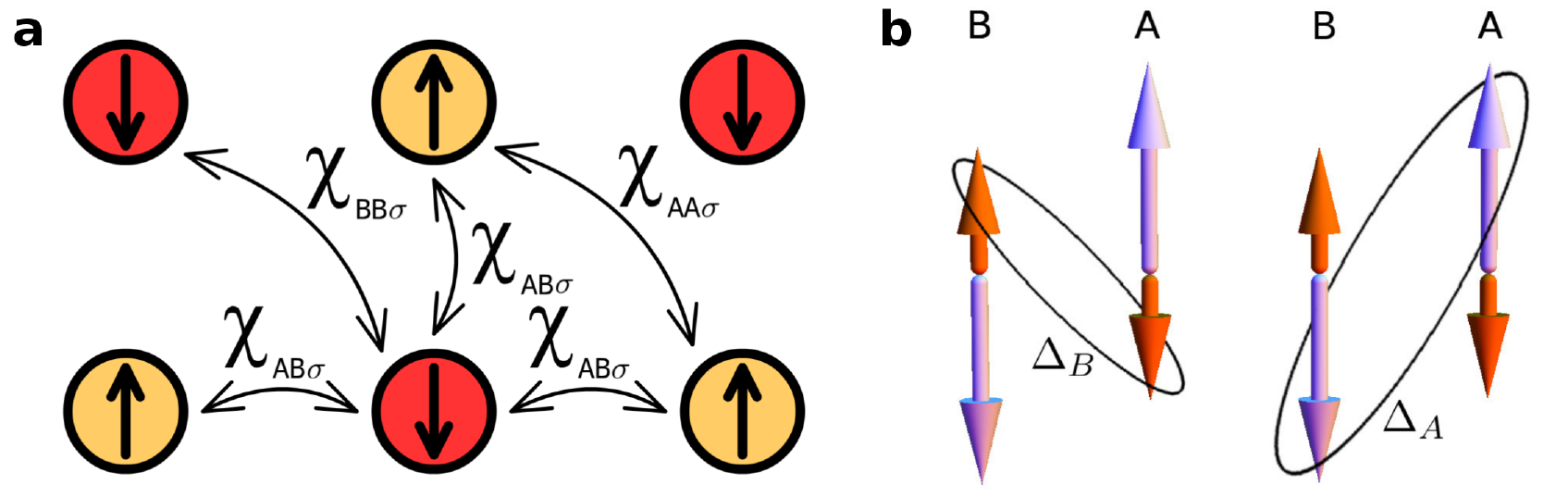}
 \caption{Schematic interpretation of $\chi_{AB\sigma} \equiv \chi_\sigma$, $\chi_{AA\sigma} $ and $\chi_{BB\sigma}$ (panel $a$) and $\Delta_A$ and $\Delta_B$ (panel $b$). To include  antiferromagnetic ordering we divide the lattice into two sublattices, $A$ and $B$ with the majority of spins being \emph{up} and \emph{down}, respectively. Thus, $\chi_{AB\sigma}$ denotes hopping of electron with spin $\sigma$ between the sublattices $A$ and $B$, while $\chi_{AA\sigma}$ and $\chi_{BB\sigma}$ denote the hopping within one sublattice ($A$ or $B$ respectively); $\Delta_A$ is the pairing amplitude between the majority spins \emph{up} from sublattice $A$ and 
 minority spins \emph{down} from $B$; $\Delta_B$ is the pairing amplitude between the majority spins \emph{up} from $B$ and minority spins \emph{down} from $A$.
 Additionally, if CDW is present, then for site belonging to the A sublattice the average (total) number of electrons is larger by $\delta_n$,
 than number of electrons on the neighboring site belonging to the B sublattice (for clarity, in this picture it was assumed that $\delta_n=0$).
 }\label{sites}
\end{figure}

In our case, the resultant expectation value is dependent on a number of quantities,
\begin{equation}
 \langle \Psi_0 | \mathcal{\hat H}_{\text{$t$--$J$--$U$--$V$}}^{\itext{eff}} | \Psi_0 \rangle  \equiv
 W(n,\, m,\, \delta_n,\, d_A,\, d_B,\, \chi,\, \chi_S,\, \chi_T,\, \Delta_S,\, \Delta_T),
\end{equation}
where $W(\ldots)$ is a functional of a number of mean-field averages
that are explained below
(for the explicit form of $W$ and the details of the calculations see ~\ref{app:GutzwillerFactors}).
First, $n$ is the average number of electrons per site, $m$ is the magnitude of staggered magnetization in AF state,
and $\delta_n$ is the order parameter for CDW phase.
Those three quantities can be combined together by expressing the local occupancy in the following manner,
\begin{equation}
 n_{i\sigma} \equiv \langle \hat c_{i\sigma}^\dagger \hat c_{i\sigma} \rangle_0 \equiv \frac{1}{2} \left( n +  e^{i {\bf Q} \cdot {\bf R}_i}\, ( \sigma m + \delta_n) \right),
 \label{eq:n_Q}
\end{equation}
where for simplicity, we denote $\langle \Psi_0 | \ldots | \Psi_0 \rangle \equiv \langle \ldots \rangle_0$.
The superlattice vector was first chosen to be ${\bf Q}=(\pi,\, \pi)$,
i.e., the lattice is naturally divided into two sublattices, A and B, such that one sublattice (A) has
in average $\frac{1}{2}(n+m+\delta_n)$ \emph{up} ($\uparrow$) electrons and
$\frac{1}{2}(n-m+\delta_n)$ \emph{down} ($\downarrow$) electrons, while
the second sublattice (B) has in average
$\frac{1}{2}(n-m-\delta_n)$ \emph{up} ($\uparrow$) and
$\frac{1}{2}(n+m-\delta_n)$ \emph{down} ($\downarrow$) electrons.
Second, the double occupancy probabilities on the sublattices are labelled as $d_A$ and $d_B$, respectively.
Third, the average hopping amplitude for the first and the next nearest neighbors (1st and 2nd n.n.) are defined by
\begin{equation}
 \chi_{ij\sigma} \equiv \langle \hat c_{i\sigma}^\dagger \hat c_{j\sigma} \rangle_0 \equiv \left\{
   \begin{array}{ll}
     \chi_{\sigma} & \hspace{3pt} \text{for 1st n.n.}, \\
     \chi_{S,\sigma} +  e^{i {\bf Q} \cdot \mathbf{R}_i} \chi_{T,\sigma} & \hspace{3pt} \text{for 2nd n.n.},
   \end{array}
\right.
\end{equation}
with $\chi_{S,\sigma} \equiv \frac{1}{2} ( \chi_{AA\sigma} + \chi_{BB\sigma} )$, $\chi_{T,\sigma} \equiv \frac{1}{2} ( \chi_{AA\sigma} - \chi_{BB\sigma})$,
where $\chi_{AA\sigma}$ and $\chi_{BB\sigma}$ denote respectively hopping of electron with the spin $\sigma$ within sublattice A and B,
and $\chi_{AB\sigma} \equiv \chi_\sigma$ is the hopping between the sublattices (cf.\ Fig.~\ref{sites} a).
Fourth, the electron pairing amplitude between nearest neighbors, with spin-singlet and triplet components,
$\Delta_S$ and $\Delta_T$, are defined by \cite{Abram2013-PhysRevB.88.094502}
\begin{equation}
 \Delta_{ij\sigma} \equiv \langle \hat c_{i \sigma} \hat c_{j \bar\sigma} \rangle_0 =
 - \tau_{ij}\left(\sigma \Delta_S + e^{i {\bf Q}\cdot {\bf R}_i} \Delta_T\right),
 \label{eq:gaps_T_S}
\end{equation}
where $ \tau_{ij} \equiv 1$ for $j=i\pm \hat x$, and $ \tau_{ij} \equiv -1$ for $j=i\pm \hat y$ to ensure the $d$-wave symmetry of $\Delta_{ij\sigma}$,
and with $\Delta_S \equiv \frac{1}{4}\left( \Delta_A + \Delta_B + \mbox{H.c.} \right)$ and $\Delta_T \equiv \frac{1}{4} \left( \Delta_A - \Delta_B + \mbox{H.c.} \right)$ (cf.\ Fig.~\ref{sites} b).

The mean-field order parameters defined above are determined numerically by minimizing the system's ground-state energy.
However, in order to be sure that the self-consistent equations for them are also fulfilled when carrying out the variational minimization procedure,
we introduce additional constraints with the help of the Lagrange multiplier method (cf. \cite{SGA,Jedrak2010,Jedrak2011}). The constraints form an essence of the statistically-consistent method, in addition to the standard Gutzwiller approximation elaborated in Appendix A.
Such an approach leads to the effective Hamiltonian of the following form
\begin{multline} 
 \hat K = W(n,\,m,\, \ldots) -
 \sum_{\langle i,j\rangle,\sigma} \left( \lambda_{ij\sigma}^\chi \left( \hat c_{i\sigma}^\dagger \hat c_{j\sigma} - \chi_{ij\sigma} \right) + \text{H.c.} \right) \\
- \sum_{\langle\langle i,j\rangle\rangle,\sigma} \left( \lambda_{ij\sigma}^{\chi} \left( \hat c_{i\sigma}^\dagger \hat c_{j\sigma} - \chi_{ij\sigma} \right) + \text{H.c.} \right) \\
- \sum_{\langle i,j\rangle} \left( \lambda_{ij\sigma}^\Delta \left( \hat c_{i\sigma} \hat c_{j\bar\sigma} - \Delta_{ij\sigma} \right) + \text{H.c.} \right) \\
- \sum_{i\sigma} \left( \lambda_{i\sigma}^n \left( \hat n_{i\sigma} - n_{i\sigma} \right) \right)
- \mu \sum_{i\sigma} \hat n_{i\sigma}.
\end{multline}
Note that the constraints introduce the renormalizations of the hoppings $\lambda_{ij\sigma}^{\chi}$ and the pairing potential $\lambda_{ij\sigma}^{\Delta}$ in real space as well as the local molecular field $\lambda_{i\sigma}^nn$ and the chemical potential shift $\lambda_{i\sigma}^n$.

Next we define the generalized grand potential function at temperature $T>0$,
 \begin{equation}
 \mathcal{F} = -\frac{1}{\beta} \ln{\mathcal{Z}}, \hspace{6pt} \mbox{with} \hspace{6pt}  \mathcal{Z} = \mathrm{Tr}\left( e^{-\beta \hat K} \right), \label{eq:Funkcjonal_SGA_def}
\end{equation}
with the Landau free energy equal to
\begin{equation}
F =  \mathcal{F}_0 +  \mu \Lambda n, \label{eq:FreeEnergy}
\end{equation}
where $\mathcal{F}_0$ denotes the value of $\mathcal{F}$ obtained at the minimum, i.e. when the following conditions are fulfilled,
\begin{equation}
\frac{\partial \mathcal{F}}{\partial A_i} = 0, \hspace{12pt}
\frac{\partial \mathcal{F}}{\partial \lambda_i} = 0, \hspace{12pt}
\frac{\partial \mathcal{F}}{\partial d_A} = 0, \hspace{12pt}
\frac{\partial \mathcal{F}}{\partial d_B} = 0,
\label{eq:set-equations}
\end{equation}
where $\{A_i\}$ denote the mean-field averages while $\{\lambda_i\}$ refers to the Lagrange multipliers.
The set of equations \bref{eq:set-equations} can be subsequently solved using numerical methods.
The results are presented in the Section \ref{sec:Results}. In the SGA approximation, we regard all the local fields and mean-field averages as spatially homogeneous, dependent only on the sublattice index. After solving the set of equations (\ref{eq:set-equations}) we can calculate the so-called correlated SC gap which is the correspondent of $\Delta_{ij\sigma}$ in the Gutzwiller state
\begin{equation}
 \Delta^G_{ij\sigma}=\langle\hat{c}_{i\sigma}\hat{c}_{j\bar{\sigma}}\rangle=\alpha_{i\sigma}\alpha_{j\bar{\sigma}}\langle\hat{c}_{i\sigma}\hat{c}_{j\bar{\sigma}}\rangle_0,
\end{equation}
where $\alpha_{i\sigma}$ is a function of $n_{i\sigma}$, $n_{i\bar{\sigma}}$, $d^2_i$ and its explicit form is given in Appendix A. The spin-singlet ($\Delta^G_S$) and spin-triplet ($\Delta^G_S$) correlated gaps are defined in the analogous way as the non-correlated ones (cf. Eq. \ref{eq:gaps_T_S}).

\subsection{Extension: DE-GWF Approach}

The SGA method described in the previous Section should be considered as a more sophisticated form of the
Renormalized Mean Field Theory (RMFT), with the statistical consistency conditions included explicitly, since the Gutzwiller approximation does not respect them \cite{Jedrak2011}.
In this Section we describe the Diagrammatic Expansion of the Gutzwiller Wave Function approach (DE-GWF)
\cite{Bunemann2012, Kaczmarczyk2013, Kaczmarczyk2014, Kaczmarczyk2015, Wysokinski2014}.
With this method one does not make use of the approximations (\ref{eq:GA}) what leads to going beyond the RMFT or SGA. This can be done in a systematic manner by including the nonlocal correlations in higher orders
and thus reach asymptotically the full Gutzwiller-wave-function solution step by step.
It is important to note that within the extended approach, the SGA is equivalent to the zeroth order form of the DE-GWF method.
As the full approach is significantly more complicated than the SGA method, here we address the
question of the full solution only for a pure superconducting phase and analyze the influence of the intersite Coulomb repulsion on that phase.
The determination of the full phase diagram, i.e., with the coexistent AF and CDW phases,
is cumbersome within DE-GWF and must be discussed separately. 

Similarly as before, in DE-GWF method we are looking for the ground state of the system in the form given by Eq. (\ref{eq:Gutzwiller_wf}). Using the condition (\ref{eq:P2_diag}), we can write all the relevant expectation values,
which appear during the evaluation of Eq. (\ref{eq:rownowaznoscRMFT_GA}), in the form of a power series
with respect to the parameter $x$ (we assume the spatial homogeneity of the paired solution, so $x\equiv x_i$), without the use of the approximations (\ref{eq:GA}). As an example, we show below the power series for the 
hopping probability and the intersite Coulomb interaction terms (all other terms remaining
in the Hamiltonian can be expressed in analogical form, cf. Ref.~\cite{Kaczmarczyk2014})

\begin{equation}
\left\{
\begin{array}{lcl}
 \langle\Psi |\hat{c}^{\dagger}_{i\sigma}\hat{c}_{j\sigma}|\Psi \rangle & = & \displaystyle \sum_{k=0}^{\infty}\frac{x^k}{k!}\sideset{}{'}\sum_{l_1\ldots l_k}\langle \tilde{c}^{\dagger}_{i\sigma}\tilde{c}_{j\sigma}\hat{d}^{\,\textrm{HF}}_{l_1\ldots l_k} \rangle_0,\vspace{6pt}\\
 \langle\Psi |\hat{n}_{i\sigma}\hat{n}_{j\sigma'}|\Psi \rangle & = & \displaystyle \sum_{k=0}^{\infty}\frac{x^k}{k!}\sideset{}{'}\sum_{l_1\ldots l_k}\langle \tilde{n}_{i\sigma}\tilde{n}_{j\sigma}\hat{d}^{\,\textrm{HF}}_{l_1\ldots l_k} \rangle_0,\\
\end{array}
\right.
\label{eq:expectation_val_terms}
\end{equation}
where $\hat{d}^{\,\textrm{HF}}_{l_1\ldots l_k}\equiv\hat{d}^{\,\textrm{HF}}_{l_1}\!\!\ldots \,\hat{d}^{\,\textrm{HF}}_{l_k}$ with $\hat{d}_{\emptyset}^{\textrm{HF}}\equiv 1$, and the primed sums have the restrictions $l_p\neq l_{p'}$ and
$l_p\neq i,j$. 
Also, the following notation has been used $\tilde{c}^{(\dagger)}_{i\sigma}=\hat{P}_i\hat{c}^{(\dagger)}_{i\sigma}\hat{P}_i$ and $\tilde{n}_{i\sigma}=\hat{P}_i\hat{n}_{i\sigma}\hat{P}_i$. By including the first 4-6 terms of the power series we are able
to calculate the expectation value of the system energy to a desired accuracy. As one can see, the inclusion of higher
order terms (i.e., those with $k>0$) leads to the situation in which the simple expressions such as, e.g.,
\begin{equation}
 \langle\Psi|\hat{c}^{\dagger}_{i\sigma}\hat{c}_{j\sigma}|\Psi\rangle = q_t \langle\Psi_0|\hat{c}^{\dagger}_{i\sigma}\hat{c}_{j\sigma}|\Psi_0\rangle,
\end{equation}
are no longer valid due to the appearance of the nonlocal correlations of increased range
(caused by the appearance of the $\hat{d}^{\,\textrm{HF}}_{l_1\ldots l_k}$ terms inside the expectation values $\langle\ldots \rangle_0$ in Eqs. (\ref{eq:expectation_val_terms})). 

By using the Wick's theorem for the averages in the non-correlated state appearing in (\ref{eq:expectation_val_terms}),
one can express the average value of the system energy in terms of the paramagnetic and superconducting
lines, i.e., the correlation functions that connects particular lattice sites
\begin{equation}
 P_{ij} \equiv \langle \hat{c}^{\dagger}_{i\sigma} \hat{c}_{j\sigma}\rangle_0, \quad S_{ij} \equiv \langle \hat{c}^{\dagger}_{i\uparrow} \hat{c}^{\dagger}_{j\downarrow}\rangle_0,
 \label{eq:lines}
\end{equation}
respectively. As we are considering only the pure SC phase in this extended approach the anomalous average corresponding to the paired state is purely of spin-singlet character. Such a procedure leads in a natural manner to the diagrammatic representation of the energy expectation value, in which the 
lattice sites play the role of the vertices of the diagrams and the paramagnetic or the superconducting
 lines are interpreted as their edges.

The minimization condition of the ground state energy \bref{eq:rownowaznoscRMFT_GA} can be evaluated by introducing
the effective single-particle Hamiltonian of the form (c.f. Ref. \cite{Spalek-arxiv})
\begin{equation}
 \hat{\mathcal{H}}_{\textrm{eff}}=\sum_{ij\sigma}t^{\textrm{eff}}_{ij}\hat{c}^{\dagger}_{i\sigma}\hat{c}_{i\sigma}
 + \sum_{ij}\big(\Delta^{\textrm{eff}}_{ij}\hat{c}^{\dagger}_{i\uparrow}\hat{c}_{i\downarrow}^\dagger  +H.c.\big),
 \label{eq:eff_Hamiltonian}
\end{equation}
where the effective parameters appearing in this Hamiltonian are defined as
\begin{equation}
 t^{\textrm{eff}}_{ij} \equiv \frac{\partial\mathcal{F}}{\partial P_{ij}},\quad \Delta^{\textrm{eff}}_{ij} \equiv \frac{\partial\mathcal{F}}{\partial S_{ij}}.
\end{equation}
By using the above concept of the effective Hamiltonian one can derive the self-consistent equations for $P_{ij}$ and $S_{ij}$,
which can then be solved numerically (cf. Ref. \cite{Kaczmarczyk2014}). Such a procedure has to be supplemented with the concomitant energy minimization
with respect to the wave-function variational parameter $x$. After determination of the value of $x$, together with those of the
paramagnetic and superconducting lines, one can evaluate the so-called correlated superconducting gap in the Gutzwiller state defined as
$\Delta_{G|ij}\equiv \langle\Psi|\hat{c}^{\dagger}_{i\uparrow}\hat{c}^{\dagger}_{j\downarrow}|\Psi\rangle/\langle\Psi|\Psi\rangle$,
which represents the physical order parameter.

It should be noted that during the calculations one may limit to the terms with lines that correspond to distances
smaller than $R_{\textrm{max}}$, as $P_{ij}$ and $S_{ij}$ with increasing distance
$|\Delta \mathbf{R}_{ij}|=|\mathbf{R}_i-\mathbf{R}_j|$ lead to systematically smaller contributions \cite{Kaczmarczyk2015}.
In our calculations we have taken $\Delta R_{max}^2=10$, which for the case of square lattice in a spatially
homogeneous state and for the $d$-wave pairing symmetry, leads to 5 different superconducting lines. 
Each of those lines has its correspondant in the correlated state. The following notation is used
in the subsequent discussion
\begin{equation}
\left\{
\begin{array}{lcl}
 \Delta^{(10)}_G & \equiv & \Delta_{G|ij}\;,\;\; \textrm{for} \;\; \Delta \mathbf{R}_{ij}=(1,0)a, \vspace{6pt}\\
 \Delta^{(20)}_G & \equiv & \Delta_{G|ij}\;,\;\; \textrm{for} \;\; \Delta \mathbf{R}_{ij}=(2,0)a, \vspace{6pt}\\
 \Delta^{(30)}_G & \equiv & \Delta_{G|ij}\;,\;\; \textrm{for} \;\; \Delta \mathbf{R}_{ij}=(3,0)a, \vspace{6pt}\\
 \Delta^{(21)}_G & \equiv & \Delta_{G|ij}\;,\;\; \textrm{for} \;\; \Delta \mathbf{R}_{ij}=(2,1)a, \vspace{6pt}\\
 \Delta^{(31)}_G & \equiv & \Delta_{G|ij}\;,\;\; \textrm{for} \;\; \Delta \mathbf{R}_{ij}=(3,1)a,
 \end{array}\right.
 \label{eq:Delta_mn}
\end{equation}
where $a$ is the lattice constant. Again, because we are now considering the pure SC phase, the gap parameters correspond to the creation of spin-singlet Cooper pairs, without the admixture of spin-triplet pairing which appears in the coexistent AF+SC phase.

%%%%%%%%%%%%%%%%%%%%%%%%%%%%%%%%%%%%%%%%%%%%%%%%%%%%%%%%%%%%%%%%%%%%%%%%%%%%%%%%%%%%%%%%%%%%%%%%%%%%%%

\section{Results}
\label{sec:Results}

\subsection{SC versus CDW stability in the Statistically Consistent Gutzwiller Approximation (SGA)}

The numerical calculations were carried out for the planar square lattice.
Unless stated otherwise, 
the following values of the microscopic parameters have been taken:
${t=-0.35}$ eV, $J=|t|/3$, $U=20|t| \approx 2.5 W$, and $\beta = 1500/|t|$,
where $\beta \equiv 1/{k_B T}$
($T$ is the absolute temperature, $k_B$ is the Boltzmann constant).
We have checked that for such a choice of $\beta$,
we reproduce accurately the $T = 0$ limit values. In the following all the energies are given in the units of $|t|$. As shown in Ref. \cite{Spalek-arxiv} similar values of $t$, $J$, and $U$ as those chosen here lead to good agreement between the DE-GWF results for the $t$-$J$-$U$ model and the princpal experimental data for the SC phase in the cuprates. The GSL library \cite{GSL-manual} has been used to solve the system up to 13 self-consistent equations.
The typical accuracy of solution was $10^{-10}$ for dimensionless quantities ($\chi_\sigma$, $\chi_{S,\sigma}$, $\chi_{T,\sigma}$, $\Delta_S$, $\Delta_T$, etc.)

\begin{figure}
 \centering
 \includegraphics[width=0.75\textwidth]{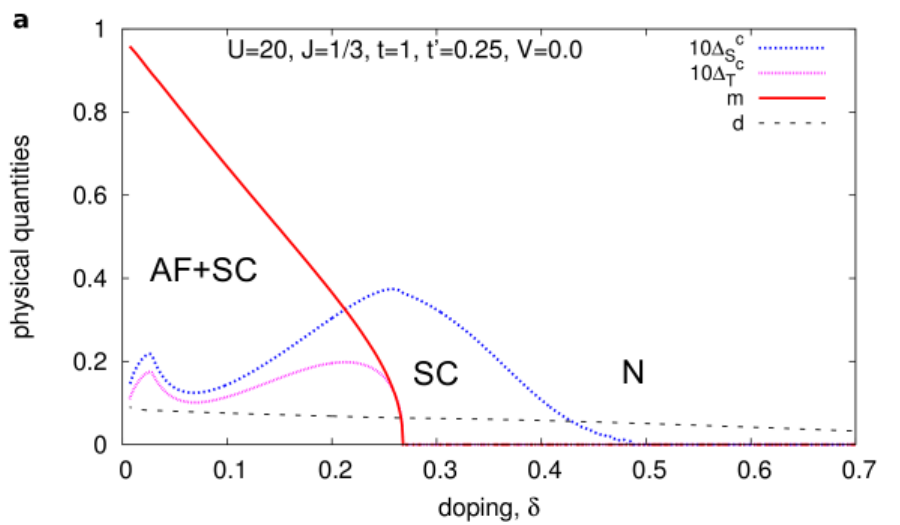}

 \includegraphics[width=0.75\textwidth]{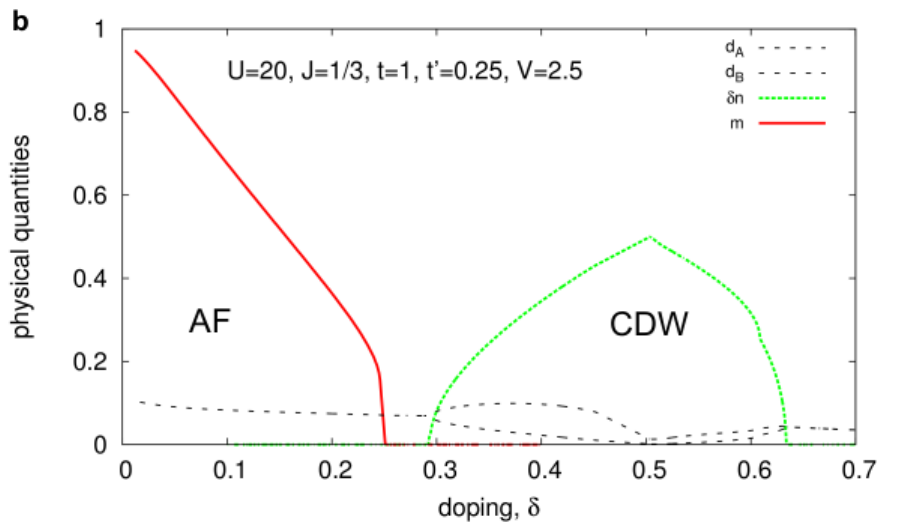}
 
 \caption{Phase diagrams containing antiferromagnetic (AF), d-wave superconducting (SC) and paramagnetic (N) phases for the two values of the intersite Coulomb interaction strength. $a)$ AF and SC phases stability regimes for $V=0$. The double occupancy only slightly changes with doping and $d^2 \approx 10^{-2}$ (note, that we plot $d$ instead of $d^2$ in this figure). $b)$ Phase diagram for $V/|t| = 2.5$. Note the appearance of the CDW stability range. However, no stable $SC$ phase for such a large value of $V$ has appeared.}
 \label{fig:CDW-V_1}
\end{figure}

\begin{figure}
 \centering
 \includegraphics[width=0.75\textwidth]{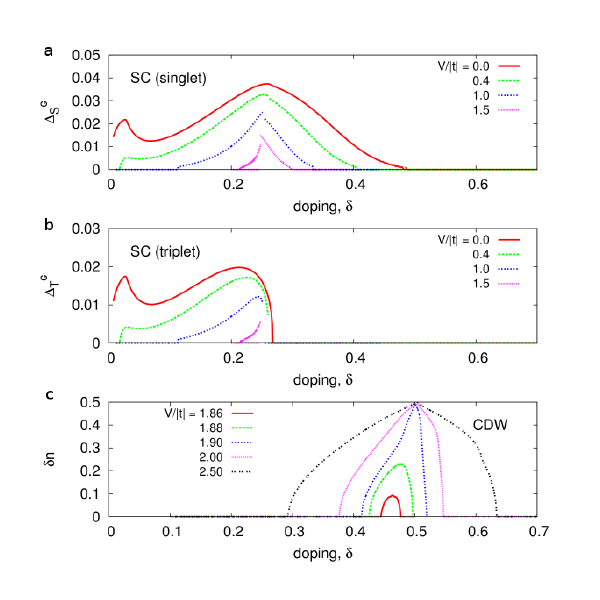}

 \caption{$a)$ and $b)$ superconducting order-parameter components $\Delta_S^G$ and $\Delta_T^G$, respectively, for selected values of $V/|t|$.
 $c)$ CDW order parameter as a function of doping for selected values of $V/|t|$. The $\Delta^G_S$ and $\Delta^G_T$ label the spin-singlet and spin-triplet pairing amplitude.}
 \label{fig:CDW-V_2}
\end{figure}

In Figure~\ref{fig:CDW-V_1}, we present the order parameters of the phases: AF, CDW, and SC,
as well as the double occupancy, $d$ (or $d_A$ and $d_B$ where necessary),
all  as a function of doping, $\delta \equiv 1-n$.
The staggered magnetization, $m$, and the difference between the average number of electrons for sublattices A and B, $\delta_n$, are the order parameters for the AF and CDW phases, respectively. In case of SC, there are two order parameters (cf. \cite{Tsonis2008-jpcm, Aperis2010-prl}), namely the singlet $\Delta_S$ and the triplet $\Delta_T$ gap parameters.
Note, that $\Delta_T \neq 0$ only if $m \neq 0$ and $d$ splits to $d_A \neq d_B$ only if $\delta_n \neq 0$
(indices $A$ and $B$ refer to different sublattices).

In Figure~\ref{fig:CDW-V_1}, we display the phase diagrams for two values of the intersite Coulomb interaction term ($V=0$ and $V=2.5$). For $V=0$ the SC phase appears for $\delta\lesssim 0.45$ with the coexisting AF+SC phase for $\delta\lesssim 0.25$ in which both the spin-singlet and the spin-triplet contributions to the pairing are present. As one can see, for relatively large value of $V=2.5$ the SC order no longer appears on the phase diagram, while the CDW phase stability region is broad. Note that the AF region is barely affected by the change of the $V$ value. 

In Fig. \ref{fig:CDW-V_2} $a$ and $b$ we show the effect of $V$ on the SC order parameters as we increase it gradually, from $0$ to $1.5$. One can see, that with the increasing value of $V$ the SC order parameters are suppressed and in the highly
underdoped region only a pure AF phase survives. In Figure~\ref{fig:CDW-V_2} $c$, the range of the CDW ordering (with the order parameter - $\delta_n$) is specified. For $V < 1.85$,
no stable CDW phase is observed, but when $V$ reaches the critical value around $1.85$, a region of a stable CDW
order appears at $\delta \approx 0.47$. Upon increasing further the value of $V$, the CDW phase regime broadens up.
Note, that here both AF and CDW states have the same modulation vector ${\bf Q} = (\pi,\, \pi)$.

It is foreseeable that for a large enough value of $V$ the stability region of the CDW phase should be centered around $\delta = n = 0.5$ on the phase diagram. This is because for this particular value of $\delta$  a given occupied atomic site is always surrounded by empty sites in the CDW phase with ${\bf Q} = (\pi,\, \pi)$ for $\delta_n=n$. As a result, the gain of the system energy resulting from the intersite Coulomb repulsion is reduced drastically. When it comes to the influence of the intersite Coulomb repulsion on the pairing, with increasing $V$ the upper concentration
of the superconductivity disappearance is reduced (from $\delta = 0.45$ to $0.3$, as observed in HTS), as well as the paired phase is destroyed in the higly underdoped region ($\delta < 0.1$). In effect, the pure AF phase becomes stable very close to half-filling. Unfortunately, for the considered modulation vector ${\bf Q} = (\pi,\, \pi)$ higher values of $V$ are required to obtain the stability of the CDW phase, which means that the observed experimentally phase diagram cannot be reproduced. Moreover, in the presented results the CDW phase appears for relatively large values of dopings. Thus, it is necessary to investigate the problem further by taking a CDW modulation vector similar to the one reported in the experiment. Such an analysis is carried out in the next Subsection. 

\subsection{CDW vs. AF stability: ${\bf Q}_{CDW}=(\frac{2}{3}\pi,\, 0)$ modulation vector}

\begin{figure}
 \centering
 \includegraphics[width=0.7\textwidth]{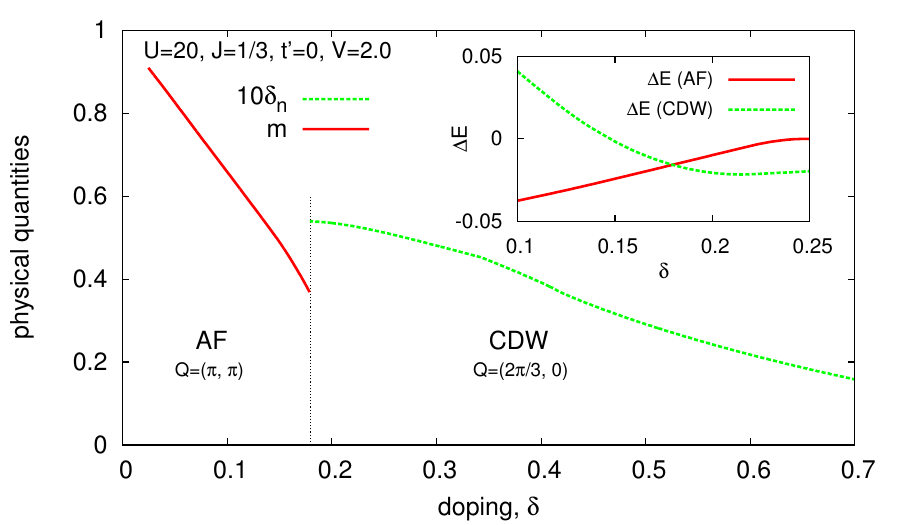}
 \caption{Phase diagram containing solely AF and CDW orders. The AF modulation vector is ${\bf Q}_{AF}=(\pi,\, \pi)$, while that corresponding to the CDW is ${\bf Q}_{CDW}=(\frac{2}{3}\pi,\, 0)$.
 Such a form of CDW modulation vector is close to that observed in the experiment \cite{Chang2012-Nature, Gerber2015-Science}. Neither SC order nor hopping between the second neighboring
 sites ($t'=0$) were included here.
 In the inset, the difference between the energies of the two solutions and the normal (paramagnetic) is presented. Note,
 that the energy of AF and CDW solution intersect, indicating the appearance of the first order transition between the two phases.}
 \label{fig:CDW-23}
\end{figure}

Here, we discuss the effect of choice of the CDW modulation vector. Namely, so far
we have assumed the simplest form of the modulation vector ${\bf Q}=(\pi,\, \pi)$, whereas 
the modulation vector reported in experiments is closer to ${\bf Q}_{CDW}=(\frac{2}{3}\pi,\, 0)$ \cite{Chang2012-Nature, Gerber2015-Science}.
Such, more realistic situation is considered next, without including the SC phase though, as the calculation of all the considered phases is quite cumbersome and should be discussed separately. By applying Eq. (\ref{eq:n_Q}) with ${\bf Q}_{CDW}=(\frac{2}{3}\pi,\, 0)$ and $m=0$ (no magnetic ordering) one can see that in the considered scenario the average number of particles per spin on an atomic site ($n_{i\sigma}$) changes in the $x$ direction with a period of $3a$ where $a$ is the lattice constant. Here, we also consider $\delta_n$ from Eq. (\ref{eq:n_Q}) as an order parameter. 
From the results depicted in Fig.\ \ref{fig:CDW-23} we see, that in this case the maximum of the CDW order parameter
is shifted towards the smaller dopings with respect to the case considered previously.
This is an important result, since it is observed in experiments that the CDW appears in the underdoped regime,
close to the boundary of AF phase \cite{Chang2012-Nature, Hucker2014-PhysRevB.90.054514}. This in turn suggests, that the full description including all phases (and their possible coexistence),
with such a choice of the CDW modulation vector might bring the theory closer to experiment.
Such a study may constitute a firm test of the one-band model applicability to HTS. It should be noted that the AF solution presented in Fig. \ref{fig:CDW-23} appears still in the too wide range of the doping $\delta$.

\subsection{SC stability beyond SGA: Effect of ntersite Coulomb interaction}
\label{sec:4.3}

\begin{figure}
 \centering
 \includegraphics[width=0.9\textwidth]{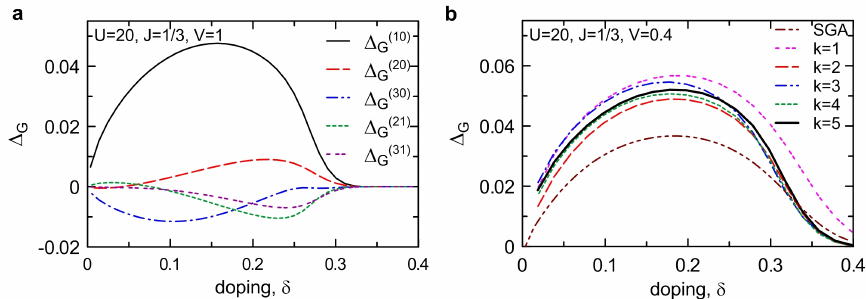}
 \caption{$a)$ Magnitudes of correlated gaps $\Delta_G^{(mn)}$ (cf. Eq.~\bref{eq:Delta_mn})
 between different neighbors as a
 function of doping for a selected set of microscopic parameters. $b)$~Evolution of the nearest-neighbor
 correlated gap $\Delta_G^{10}$ with the increasing order $k$ of computations.
 The SGA results (the lowest line)
 are equivalent to the zeroth order DE-GWF approach.
Note, that the lower critical concentration of the SC order is shown here to be $0$ even for $V>0$.
It results from the circumstance that we did not include the AF order in this graph.
For $V>0$ the AF order competes with SC in the underdoped region and a pure AF phase wins over, as shown in Fig.~\ref{fig:CDW-V_1}.}
 \label{fig:orders_DEGWF}
\end{figure}

For the sake of completeness, in this subsection we discuss the robustness of the pure superconducting phase within the 
DE-GWF method, with the intersite Coulomb repulsion included, when going beyond SGA (of RMFT type) approach.
First, we show the differences between the SGA and the DE-GWF for the selected 
set of the model parameters. The magnitudes of the intersite correlated gap parameters (cf. Eq. (\ref{eq:Delta_mn})) in an exemplary situation, obtained in the diagrammatic approach
 are displayed in Fig.\ \ref{fig:orders_DEGWF} $a$. It should be noted that for the case of pure SC phase we have only the singlet SC gap wich we denote $\Delta_G\equiv\Delta^G_S$ here. As one can see, the nearest neighbor pairing amplitude $\Delta^{(10)}_G$ 
is by far the dominant one. Nonetheless, the remaining larger-distance contributions, may also become significant.
Note, that in some doping regions different contributions can change their signs.
For example, in the underdoped regime both $\Delta^{(10)}_{G}$ and $\Delta^{(30)}_{G}$ obey exactly the $d$-wave symmetry,
but with the opposite signs. The situation is different within the SGA method, where the only nonzero pairing contribution taken into account is the nearest neighboring one.
In Fig.\ \ref{fig:orders_DEGWF} $b$ we present the evolution of the $\Delta^{(10)}_{G}$ gap with the increasing order of the calculations. The lowest dotted--dashed line corresponds to the SGA method which is also equivalent to the zeroth order of the DE-GWF approach.
The differences between the green dotted line (the fourth order) and the black solid line (the fifth order) are very small
which means that we have achieved a convergence with the assumed accuracy.
As one can see, the two methods, SGA and DE-GWF, are qualitatively similar when it comes to the doping dependence of the correlated gap, but the correlations increase the
pairing amplitude by $30\%$--$40\%$ in the latter case as it encompasses also the more distant pairing amplitudes. It should be noted that qualitative differences between the two methods appear when it comes to the appearance of the non-BCS regime as discussed in the remaining part of this work. Also, more detailed comparison of DE-GWF and SGA is presented in Ref. \cite{Spalek-arxiv} for the case of the $t$-$J$-$U$ model.

%%%%%%%%%%%%%%%%%%%%%%%%%%%%%%%%%%%%%%%%%%%%%%%%%%%%%%%%%%%%%%%%%%%%%%%%%%%%%%%%%%%%%%%%%%
%%%%%%%%%%%%%%%%%%%%%%%%%%%%%%%%%%%%%%%%%%%%%%%%%%%%%%%%%%%%%%%%%%%%%%%%%%%%%%%%%%%%%%%%%%

\begin{figure}
 \centering
 \includegraphics[width=0.9\textwidth]{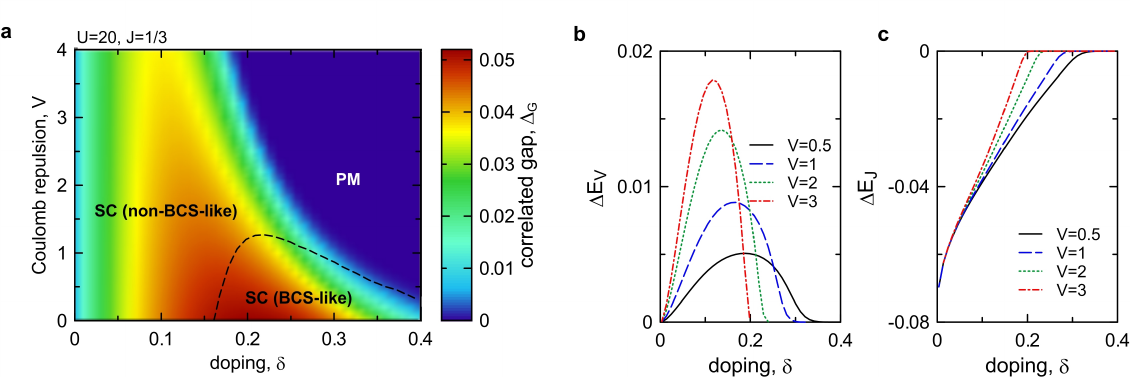}
 \caption{$a)$ Correlated gap value as a function of both doping ($\delta$) and the intersite Coulomb 
 repulsion ($V$). 
 One can distinguish between the two superconducting regimes: BCS-like and non-BCS, defined
 in the main text. $b)$ and $c)$ Contribution to the condensation energy coming from the 
 intersite Coulomb term $\sim V$ and the exchange term $\sim J$, respectively. 
 While the Coulomb repulsion $\sim V$ increases the energy of SC with respect to the 
 normal (PM) state, the opposite is true for the exchange part. 
 In this sense, the Coulomb interactions $\sim V$ play a destructive role for the pairing,
 whereas those $\sim J$ promotes the paired state, together with the kinetic-energy gain in the underdoped regime.
 }
 \label{fig:diagram_DEGWF}
\end{figure}

In Figure \ref{fig:diagram_DEGWF} we illustrate the influence of the intersite Coulomb repulsion on the 
stability of the paired phase within the DE-GWF method. As one can see, the upper critical doping
for the disappearance of the superconducting phase decreases significantly when the $V$ term
is included, similarly to the case of the SGA calculations. One of the differences between the two considered methods is that
the inclusion of the higher order contributions leads to the appearance
of the so-called non-BCS region in the considered model which is manifested by the kinetic energy gain 
at the transition to the SC phase. The kinetic energy gain is defined by
\begin{equation}
 \Delta{E}_{\textrm{kin}}\equiv E^{SC}_{G}-E^{PM}_{G},\quad E_{G}\equiv \frac{1}{N}\sideset{}{'}\sum_{ij\sigma}t_{ij}\langle\hat{c}^{\dagger}_{i\sigma}\hat{c}_{j\sigma}\rangle_G,
\end{equation} 
where $E^{SC}_{G}$ and $E^{PM}_{G}$ correspond to the kinetic energies in SC and normal (paramagnetic, PM)
phases, respectively. In the BCS-like region $\Delta{E}_{\textrm{kin}}>0$, which is also
true for the BCS theory of the phonon-mediated superconductivity, whereas
$\Delta{E}_{\textrm{kin}}<0$ for the non-BCS region.
% The appearance of the non-BCS regime is purely attributed to the higher-order correlation effects,
% taken into account within the DE-GWF method, and cannot be accounted for within SGA or any other form of RMFT.
It should be noted that the non-BCS behavior has been detected experimentally 
\cite{Deutscher2005,Giannetti2011} for the underdoped samples of the cuprate compounds. This very feature speaks for the necessity of including
the higher orders to describe this important aspects of 
cuprate superconductivity.
We show here also that
the intersite Coulomb repulsion promotes the non-BCS behavior by pushing it to higher doping values 
(cf. Fig. \ref{fig:diagram_DEGWF} a).
Therefore, even though the intersite Coulomb interaction has a destructive effect by diminishing the condensation energy
(cf. Fig.~\ref{fig:diagram_DEGWF} $b)$, it extends the region of the non-BCS state at the same time.
In Figures \ref{fig:diagram_DEGWF} $b$ and $c$ we plot explicitly the contributions to the condensation energy that originates from
either the intersite Coulomb repulsion term (b) or the exchange interaction term (c), respectively.
The intersite Coulomb repulsion term increases the energy of SC phase
with respect to the normal (PM) state which means that it has a negative influence on the pairing strength. 
The opposite is true for the case of the exchange term.

%%%%%%%%%%%%%%%%%%%%%%%%%%%%%%%%%%%%%%%%%%%%%%%%%%%%%%%%%%%%%%%%%%%%%%%%%%%%%%%%%%%%%%%%%%
%%%%%%%%%%%%%%%%%%%%%%%%%%%%%%%%%%%%%%%%%%%%%%%%%%%%%%%%%%%%%%%%%%%%%%%%%%%%%%%%%%%%%%%%%

%%%%%%%%%%%%%%%%%%%%%%%%%%%%%%%%%%%%%%%%%%%%%%%%%%%%%%%%%%%%%%%%%%%%%%%%%%%%%%%%%%%%%%%%%%%%%%%%%%%%%%

\section{A critical overview}

In the first part we have analyzed the stability of AF, SC, CDW (and some of the possible coexistent phases) within the $t$--$J$--$U$--$V$ model.
For this purpose we have used both the SGA and the DE-GWF methods.
By using the former approach we have shown that CDW phase with ${\bf Q} = (\pi,\, \pi)$ is stable only above a critical value of $V$
(the intersite Coulomb repulsion), and is detrimental to SC phase stability.
With the increasing $V$, the CDW stability range broadens up, whereas the SC phase is gradually suppressed.
Such a behavior is consistent with the experimental findings, according to which CDW and SC compete with each other. However, it should be mentioned that according to our calculations the CDW $(\pi,\, \pi)$ phase becomes stable in the overdoped regime, with maximal order parameter for $\delta\approx 0.5$, while in the experiment this phase is observed for smaller doping values. By changing the modulation vector to ${\bf Q} = (\frac{2}{3}\pi,\, 0)$, which is close to the one detected experimentally, we have obtained a shift of the CDW order parameter maximum towards the smaller dopings. Nevertheless, the region of stability of CDW is still too wide, even for
moderate values of $V$. An interesting question arises whether such a situation could change
if we included in Fig.~\ref{fig:CDW-23} the SC phase and/or went beyond the RMFT approach.

In the second part of the article, we have analyzed first
the influence of the higher-order terms on the pure SC phase
within the DE-GWF method. This was motivated by the question to what extent the former results, obtained within the $t$-$J$-$U$ model \cite{Spalek-arxiv, Spalek1977-PhysicsBC} are robust in the presence of the intersite Coulomb interaction.
One of the differences between the DE-GWF and SGA is that in the former approach
the larger-distance contributions to the pairing in real space appear (cf.\ Fig.\ \ref{fig:orders_DEGWF}), whereas
in SGA only the nearest neighbor SC gap appears. Secondly, 
the magnitude of the SC order parameter in DE-GWF is enhanced by up to $40\%$ as compared to
the zeroth-order calculations (SGA). Furthermore, the non-BCS region shows up only after the inclusion of the higher-order terms within the DE-GWF and the $V$ term pushes it to higher dopings (cf. Fig. \ref{fig:diagram_DEGWF}). The observation of the non-BCS behavior in the experiment for the cuprates \cite{Deutscher2005, Giannetti2011, Molegraaf2239, PhysRevB.74.064510} clearly shows the necessity of including the higher-order terms in the calculations for the considered model.

In experiment, it is observed that CDW and SC phases can coexist \cite{Keimer2015-Nature, Comin2014-Science, Ghiringhelli2012-Science, Chang2012-Nature, daSilva2014-Science, Hucker2014-PhysRevB.90.054514, Canosa2014-PhysRevB.90.054513}. However, in SGA the increase of the $V$ parameter
suppresses SC before CDW appears. Therefore, the inclusion of the higher-order terms, which also results in the increase of the pairing amplitude, can lead to the stability of the coexistent SC-CDW phase. The DE-GWF calculations for the CDW phase within the one-band approach, even though significantly more complicated, can still be possible to carry out. We should see progress along this line in the near future. Moreover, it would be interesting to explore the possibility of the intraunitcell charge ordering appearance within the 3-band Emery type or related model by using the same method \cite{Comin2015-nature}. The combined exact diagonalization-ab-initio approach \cite{Kadzielawa2015-PRB,Biborski2015-CCP} for the Cu-O clusters could also bring some new insights in the field of charge ordered states of the cuprates.

%%%%%%%%%%%%%%%%%%%%%%%%%%%%%%%%%%%%%%%%%%%%%%%%%%%%%%%%%%%%%%%%%%%%%%%%%%%%%%%%%%%%%%%%%%%%%%%%%%%%%%

\section*{Acknowledgements}
JS and MA acknowledge the financial support through the Grant MAESTRO, No. DEC-2012/04/A/ST3/00342 from the National Science Centre (NCN) of Poland. MZ acknowledges the financial support through the Grant SONATA No. 2016/21/D/ST3/00979 from the National Science Centre, Poland.

\appendix
\section{Form of the $W$ function and mean-field renormalization factors}
\label{app:GutzwillerFactors}

The central problem within the SGA approach is to calculate the expectation value of the system energy $\langle \psi\, |\, \mathcal{\hat H}_{\text{$t$--$J$--$U$--$V$}}\, |\, \psi \rangle$ (cf.\ Eq.~\ref{eq:rownowaznoscRMFT_GA}), namely
\begin{multline}
 W = \langle  \mathcal{\hat H}_{\text{$t$--$J$--$U$--$V$}}\rangle =
 t \sum_{\langle i,\, j \rangle,\, \sigma} \left( \langle \hat c^\dagger_{i\sigma} \hat c_{j\sigma} \rangle + \mathnormal{H.c.} \right) \\
 + t' \sum_{\langle\langle i,\, j \rangle\rangle,\, \sigma} \left( \langle \hat c^\dagger_{i\sigma} \hat c_{j\sigma} \rangle + \mathnormal{H.c.} \right)
 + J \sum_{\langle i,\, j \rangle} \langle \hat {\bf S}_i \cdot \hat {\bf S}_j \rangle \\
 + U \sum_{i} \langle \hat n_{i\uparrow} \hat n_{i\downarrow} \rangle
 + \underbrace{\left( \tilde V - \frac{1}{4} J \right)}_{V} \sum_{\langle i,\, j \rangle, \sigma, \sigma'}\!\!\!\! \langle \hat n_{i\sigma} \hat n_{j\sigma'} \rangle,
\label{eq:W-averages}
 \end{multline}
where for simplicity we denote $\langle \Psi | \ldots | \Psi \rangle \equiv \langle \ldots \rangle$ and 
the wave function is taken in the form $| \Psi \rangle = \hat P\, | \Psi_0 \rangle$\cite{Gutzwiller1962, Gutzwiller1965}, where $| \Psi_0 \rangle$ is a simple,
non-correlated wave function, and $\hat P = \prod_i \hat P_i$ is the operator, which
changes the probability of local states appearing at atomic sites. The most general form of $\hat{P}_i$ is
\begin{multline}
 \hat P_i = \sum_{\Gamma} \lambda_{\Gamma} | \Gamma\rangle_{i\,i} \langle \Gamma| =
 \lambda_{i,0} (1 - \hat n_{i\uparrow}) (1 - \hat n_{i\downarrow}) + \\
 \lambda_{i,\uparrow} \hat n_{i\uparrow} (1 - \hat n_{i\downarrow}) +
 \lambda_{i,\downarrow} (1 - \hat n_{i\uparrow}) \hat n_{i\downarrow} +
 \lambda_{i,d} \hat n_{i\uparrow} \hat n_{i\downarrow}.
 \label{eq:P-general}
\end{multline}
Following Ref.~\cite{Bunemann2012}, we assume that $\hat P_i^2 \equiv 1 + x_i\, \hat n_{i\uparrow}^{HF} \hat n_{i\downarrow}^{HF}$,
with $\hat n_{i\sigma}^{HF} \equiv \hat n_{i\sigma} - n_{i\sigma}$.
Next, by acting with $\hat P_i^2$ on the states from the local basis, 
$|\emptyset\rangle_i$, $|\!\!\uparrow\rangle_i$, $|\!\!\downarrow\rangle_i$, $|\!\!\uparrow\downarrow\rangle_i$),
one yields:
\begin{equation}
\left\{
\begin{array}{lcl}
 \lambda_{i,0}^2 & = & 1 + x_i\, n_{i\sigma} n_{i\bar\sigma}, \vspace{6pt}\\
 \lambda_{i,\sigma}^2 & = & 1 - x_i\, (1 - n_{i\sigma}) n_{i\bar\sigma}, \vspace{6pt}\\
 \lambda_{i,d}^2 & = & 1 + x_i\, (1 - n_{i\sigma}) (1 - n_{i\bar\sigma}),
\end{array}
\right.
\label{eq:lambdas}
\end{equation}
where $x_i$ is a variational parameter.
When $\forall_i\, x_i=0$, then the operator $\hat P = \mathbb{I}$ and $|\Psi \rangle = |\Psi_0 \rangle$,
but when $\exists_i\, x_i<0$, then the probability of double occupancy on site $i$ is reduced.
Since the average number of electrons in the system should remain constant,
$x_i<0$ requires, that the number of the single occupied sites is increased and the number of
empty sites is reduced at the same time. 

A direct meaning of parameter $x_i$ is not easy to provide. Therefore we introduce
$d_i^2$ as the double-occupancy probability at site $i$, namely, 
\begin{equation}
 \langle \Psi | \hat n_{i\uparrow} \hat n_{i\downarrow} | \Psi \rangle \equiv d_i^2.
 \label{eq:bindingx_1}
\end{equation}
We can relate $d_i^2$ to the $x_i$ parameter, since
\begin{equation}
 d_i^2 = \langle \Psi | \hat n_{i\uparrow} \hat n_{i\downarrow} | \Psi \rangle = \langle \Psi_0 | \hat P_i \hat n_{i\uparrow} \hat n_{i\downarrow} \hat P_i | \Psi_0 \rangle
 = \lambda_{id}^2 n_{i\uparrow} n_{i\downarrow},
  \label{eq:bindingx_2}
\end{equation}
where we have assumed that $\langle \Psi_0 | \hat n_{i\uparrow} \hat n_{i\downarrow} | \Psi_0 \rangle \equiv \langle \hat n_{i\uparrow} \hat n_{i\downarrow} \rangle_0 = n_{i\uparrow} n_{i\downarrow}$,
i.e., that the following averages $\langle \hat c_{i\uparrow} \hat c_{i\downarrow} \rangle_0$
and $\langle \hat c_{i\uparrow}^\dagger \hat c_{i\downarrow} \rangle_0$ are zero.
Using Eqs.~\bref{eq:lambdas}--\bref{eq:bindingx_2}, we can show, that
\begin{equation}
 x_i \equiv \frac{d_i^2 - n_{i\uparrow} n_{i\downarrow}}{n_{i\uparrow} n_{i\downarrow} (1 - n_{i\uparrow}) (1 - n_{i\downarrow})},
\end{equation}
and as a result, we can rewrite the expressions \bref{eq:lambdas} in the form:
\begin{eqnarray}
 \lambda_{i0}^2 & = & \frac{1+ d_i^2 - n_{\sigma} - n_{\bar\sigma}}{(1 - n_{\sigma}) (1 - n_{\bar\sigma})}, \\
 \lambda_{i\sigma}^2 & = & \frac{n_{\sigma} - d_i^2}{n_{\sigma} (1 - n_{\bar\sigma})}, \\
 \lambda_{id}^2 & = & \frac{d_i^2}{n_{\sigma} n_{\bar\sigma}}. 
\end{eqnarray}

To calculate the averages appearing in Eq.~\bref{eq:W-averages}, we need one more (partial) result, namely
\begin{equation}
\begin{split}
  \hat P_i \hat c_{i\sigma}^\dagger \hat P_i & = 
 \left( \lambda_\sigma \hat n_{i\sigma} (1 - \hat n_{i\bar\sigma}) + \lambda_d \hat n_{i\sigma} \hat n_{i\bar\sigma} \right) \hat c_{i\sigma}^\dagger
  \left( \lambda_{\bar\sigma} \hat n_{i\bar\sigma} (1 - \hat n_{i\sigma}) + \lambda_0 (1 - \hat n_{i\sigma}) (1 - \hat n_{i\bar\sigma}) \right)\\
%  & = \left( \lambda_\sigma \lambda_0 + (\lambda_d \lambda_{\bar\sigma} - \lambda_\sigma \lambda_0) n_{\bar\sigma} \right) \hat c_{i\sigma}^\dagger + 
%  \left( \lambda_d \lambda_{\bar\sigma} - \lambda_\sigma \lambda_0 \right) \hat n_{i\bar\sigma}^{HF} \hat c_{i\sigma}^\dagger \\
 &= (\alpha_{i\sigma} + \beta_{i\sigma} \hat n_{i\bar\sigma}^{HF} ) \hat c_{i\sigma}^\dagger,
 \label{eq:PfP}
\end{split}
\end{equation}
where
\begin{eqnarray}\allowdisplaybreaks
 \alpha_{i\sigma} & = & \sqrt{\frac{(n_{i\sigma} - d_i^2)(1-n+d_i^2)}{n_{i\sigma}(1 - n_{i\sigma})}} + |d_i| \sqrt{\frac{n_{i\bar\sigma} - d_i^2}{n_{i\sigma}(1-n_{i\sigma})}},\\
 \beta_{i\sigma} & = & -\sqrt{\frac{(n_{i\sigma} - d_i^2)(1-n+d_i^2)}{n_{i\sigma}(1 - n_{i\sigma})(1-n_{i\bar\sigma})^2}} + |d_i| \sqrt{\frac{n_{i\bar\sigma} - d_i^2}{n_{\sigma}n_{\bar\sigma}^2(1-n_{\sigma})}}.
\end{eqnarray} 

Note that for $\hat P_i \hat c_{i\sigma} \hat P_i$ we would obtain the same result as above.
Using the obtained expressions, one can calculate other averages,
e.g., the average of the hopping term is then
\begin{multline}
 \langle \hat c_{i\sigma}^\dagger \hat c_{j\sigma} \rangle 
=  \langle \hat P_i \hat P_j \hat c_{i\sigma}^\dagger \hat c_{j\sigma} \hat P_i \hat P_j \rangle_0 
=  \langle \hat P_i  \hat c_{i\sigma}^\dagger\hat P_i \, \hat P_j \hat c_{j\sigma}  \hat P_j \rangle_0  \\
=  \alpha_{i\sigma}\alpha_{j\sigma} \langle \hat c_{i\sigma}^\dagger \hat c_{j\sigma}\rangle_0
+  \alpha_{i\sigma}\beta_{j\sigma} \langle \hat n_{i\bar\sigma}^{HF} \hat c_{i\sigma}^\dagger \hat c_{j\sigma}\rangle_0 \\
+  \alpha_{j\sigma}\beta_{i\sigma} \langle \hat n_{j\bar\sigma}^{HF} \hat c_{i\sigma}^\dagger \hat c_{j\sigma}\rangle_0 
+  \beta_{i\sigma}\beta_{j\sigma} \langle \hat n_{i\bar\sigma}^{HF} \hat n_{j\bar\sigma}^{HF} \hat c_{i\sigma}^\dagger \hat c_{j\sigma}\rangle_0.
\label{eq:hoppin-derivation-q}
\end{multline}
Using the Wick's theorem we can check that $\langle \hat n_{i\bar\sigma}^{HF} \hat c_{i\sigma}^\dagger \hat c_{j\sigma}\rangle_0 = 0$ and
$\alpha_{j\sigma}\beta_{i\sigma} \langle \hat n_{j\bar\sigma}^{HF} \hat c_{i\sigma}^\dagger \hat c_{j\sigma}\rangle_0 = 0$,
as far as we assume that there is no onsite pairing of electrons,
${\langle \hat c_{i\sigma}^\dagger \hat c_{i\bar\sigma}^\dagger \rangle_0 = 0}$, and
hopping does not change spin,
${\langle \hat c_{i\sigma}^\dagger \hat c_{j\bar\sigma} \rangle_0 = 0}$.
The last average in Eq.~\ref{eq:hoppin-derivation-q} is usually non-zero, but small, therefore it can be neglected here.
Hence, we are left with
\begin{equation}
 \langle \hat c_{i\sigma}^\dagger \hat c_{j\sigma} \rangle \approx \alpha_{i\sigma}\alpha_{j\sigma} \langle \hat c_{i\sigma}^\dagger \hat c_{j\sigma}\rangle_0.
\end{equation}
In the simplest case, where neither AF nor CDW orderings are present, we have $\alpha_{i\sigma} = \alpha_{j\sigma} = \alpha$ and thus
\begin{equation}
 \alpha^2 = g_t \equiv \frac{n - 2d^2}{n(1-n/2)} \left( \sqrt{1-n+d^2} + |d| \right)^2,
 \label{eq:gt-noAFnoCDW}
\end{equation}
which is the Gutzwiller factor for the hopping part, well known from the literature, cf. Refs.~\cite{Zhang1988-PhysRevB.37.3759, Ogawa1975-ProgTheorPhys.53.614, Vollhardt1984-rmp, Ogata2008-RepProgPhys}.

In a similar manner, the other averages appearing in Eq.~\bref{eq:W-averages} can also be calculated.
To provide one more example, we show here explicitly how to calculate the last term,
$\langle \hat n_{i\sigma} \hat n_{j\sigma'} \rangle$,
requiring perhaps the most non-trivial calculations rarely discussed in the literature.
To abbreviate the length of the expressions,
we assume here for a moment, that we are interested only in the AF order.
The generalization to the case of CDW order (or others) is not difficult and it can be left to the Reader.
Note that in the simplest two-sublattice AF ordering it is required that $n_\sigma$ on A sublattice is equal to $n_{\bar\sigma}$ on the B sublattice.
In such a case,
\begin{equation}\allowdisplaybreaks
\begin{split}
 \Lambda^{-1} & \sum_{\langle i,j \rangle,\, \sigma,\, \sigma'} \langle \hat n_{i\sigma} \hat n_{j\sigma'} \rangle  = 
 \Lambda^{-1} \sum_{\langle i,j \rangle,\, \sigma,\, \sigma'} \langle \hat P_i \hat n_{i\sigma} \hat P_i \hat P_j \hat n_{j\sigma'} \hat P_j\rangle_0 = \\
 & = \Lambda^{-1} \sum_{\langle i,j \rangle,\, \sigma,\, \sigma'} \left\langle
 \left( \hat n_{i\sigma} + (\lambda_d^2 - \lambda_{\sigma}^2) \hat n_{i\sigma} \hat n_{i\bar\sigma}^{HF} \right)
 \left( \hat n_{j\sigma'} + (\lambda_d^2 - \lambda_{\bar\sigma'}^2) \hat n_{j\sigma'} \hat n_{j\bar\sigma'}^{HF} \right) \right\rangle_0 \\
 & \approx \Lambda^{-1} \sum_{\langle i,j \rangle,\, \sigma}
 \langle n_{i\sigma} n_{j\sigma}\rangle_0 + \langle n_{i\sigma} n_{j\bar\sigma}\rangle_0 \\
& \hspace{24pt} + \Big(
   \langle \hat n_{j\bar\sigma} \rangle_0 \langle \hat n_{i\bar\sigma} \hat n_{j\sigma}^{HF} \rangle_0
 + \langle \hat n_{j\bar\sigma} \rangle_0 \langle \hat n_{i\sigma} \hat n_{j\sigma}^{HF} \rangle_0
 + \langle \hat n_{i\sigma} \rangle_0 \langle \hat n_{j\sigma} \hat n_{i\bar\sigma}^{HF} \rangle_0 \\
& \hspace{24pt} +
   \langle \hat n_{i\sigma} \rangle_0 \langle \hat n_{j\bar\sigma} \hat n_{i\bar\sigma}^{HF} \rangle_0
  \Big) (\lambda_d^2 - \lambda_\sigma^2) \\
& \hspace{24pt} + (\lambda_d^2 - \lambda_\sigma^2) (\lambda_d^2 - \lambda_{\bar\sigma}^2)
\langle \hat n_{i\sigma} \rangle_0 \langle \hat n_{j\sigma} \rangle_0 \langle \hat n_{i\bar\sigma}^{HF} \hat n_{j\bar\sigma}^{HF} \rangle_0\\
& \hspace{24pt} + (\lambda_d^2 - \lambda_\sigma^2) (\lambda_d^2 - \lambda_{\sigma}^2)
\langle \hat n_{i\sigma} \rangle_0 \langle \hat n_{j\bar\sigma} \rangle_0 \langle \hat n_{i\bar\sigma}^{HF} \hat n_{j\sigma}^{HF} \rangle_0 \\ 
&  = 2n^2 + (-4 \chi^2 + 4 \Delta_S^2 + 4 \Delta_T^2)
\left(1 + n_{\sigma}(\lambda_d^2 - \lambda_{\sigma}^2) + n_{\bar\sigma}(\lambda_d^2 - \lambda_{\bar\sigma}^2) \right) \\
& \hspace{24pt} + 4 n_{\sigma}(\lambda_d^2 - \lambda_{\sigma}^2)n_{\bar\sigma}(\lambda_d^2 - \lambda_{\bar\sigma}^2) (-\chi^2)\\
& \hspace{24pt} + 2\left( \left[ n_{\sigma}(\lambda_d^2 - \lambda_{\sigma}^2) \right]^2 + \left[ n_{\bar\sigma}(\lambda_d^2 - \lambda_{\bar\sigma}^2) \right]^2 \right) (\Delta_S^2 + \Delta_T^2)\\
& = 2 n^2 + 4 g_{v}^{\chi} ( -\chi^2) + 4 g_{v}^{\Delta} ( \Delta_S^2 + \Delta_T^2), \label{eq:gV-noCDW}
\end{split}
\end{equation}
where
\begin{equation}
 \begin{split}
 g_{v}^{\chi} &\equiv \left(1 + n_{\sigma}(\lambda_d^2 - \lambda_{\sigma}^2) + n_{\bar\sigma}(\lambda_d^2 - \lambda_{\bar\sigma}^2) 
 + n_{\sigma}(\lambda_d^2 - \lambda_{\sigma}^2)n_{\bar\sigma}(\lambda_d^2 - \lambda_{\bar\sigma}^2) \right),\\
 g_{v}^{\Delta} & \equiv  \left(1 + n_{\sigma}(\lambda_d^2 - \lambda_{\sigma}^2) + n_{\bar\sigma}(\lambda_d^2 - \lambda_{\bar\sigma}^2) 
 + \frac{1}{2}\left( \left[ n_{\sigma}(\lambda_d^2 - \lambda_{\sigma}^2) \right]^2 + \left[ n_{\bar\sigma}(\lambda_d^2 - \lambda_{\bar\sigma}^2) \right]^2 \right) \right).
 \end{split}
\end{equation}
The ``$\approx$'' sign in Eq.~\bref{eq:gV-noCDW} results from the fact, that we neglected terms proportional to
$\chi^4$, $\Delta_S^4$, $\Delta_T^4$ and $n_{\sigma} n_{\bar\sigma} \chi^2$.
Note that if no AF order is considered ($m=0$), then $n_{\sigma} = n_{\bar\sigma} = n/2$ and then
\begin{equation}
 g_{v}^{\chi} = g_{v}^{\Delta} = \left( \frac{2d^2 + n(1-n)}{n(1-n/2)} \right)^2.
\end{equation}

\section{Two ways of defining the Gutzwiller factor in the presence of extra orderings}
\label{app:twoWays}

In this Appendix we show that the introduction of extra orderings, such as AF or CDW,
can lead to a specific ambiguity in determining the final form of the Gutzwiller renormalization factors.
We explain also, how we have decided to select a particular form used in main text.

For simplicity, we assume here, that $U \rightarrow \infty$,
resulting in $d \rightarrow 0$. Furthermore, to make our arguments easy to follow, we consider only the AF order
and focus on the example of Gutzwiller factor for the hopping term
that has been already discussed in the foregoing Appendix (cf.\ Eq.~\bref{eq:gt-noAFnoCDW}).
The generalization to other states and to other averages is straightforward.

Within the SGA approach one expresses the average from the hopping term in the following manner
\begin{equation}
 \langle \Psi | \hat c_{i\sigma}^\dagger \hat c_{j\sigma} | \Psi \rangle
 \approx \langle \Psi_0 | \hat P_i \hat c_{i\sigma}^\dagger \hat P_i\, \hat P_j\hat c_{j\sigma} \hat P_j | \Psi_0 \rangle
%  & \approx & g_t \langle \Psi_0 | \hat c_{i\sigma}^\dagger \hat c_{j\sigma} | \Psi_0 \rangle.
\label{app:gt_def1}
\end{equation}
The renormalization factor for particular processes can be defined in the following manner
\begin{equation}
 g_t(n_{i\sigma}, n_{i\bar\sigma}, d, \ldots) = \frac{\langle \Psi | \hat c_{i\sigma}^\dagger \hat c_{j\sigma} | \Psi \rangle}{\langle \Psi_0 | \hat c_{i\sigma}^\dagger \hat c_{j\sigma} | \Psi_0 \rangle}. 
 \label{app:gt_def}
\end{equation}

Let us assume that the average number of electrons per atomic site is $n$ and the staggered magnetization is equal to $m$.
In the non-correlated case ($U=0$), there is in average $n_\sigma \equiv n_{A\sigma} = \frac{1}{2}(n+\sigma m)$ electrons with spin $\sigma$ per site
for the A sublattice and $n_{\bar\sigma} \equiv n_{B\sigma} = \frac{1}{2}(n-\sigma m)$ for the B sublattice.
Additionally, in average,
$n_{\uparrow\downarrow} = n_{A\uparrow} n_{A\downarrow} = n_{B\uparrow} n_{B\downarrow}$ and
consequently, $n_{\varnothing} = (1-n_{A\uparrow}) (1-n_{A\downarrow}) = (1-n_{B\uparrow}) (1-n_{B\downarrow})$
(cf. Table~\ref{tab:Psi0}).

\begin{table}
\caption{Likelihood of a site being in the certain state (uncorrelated case $|\Psi_0 \rangle$).}\label{tab:Psi0}
\centering
\begin{tabular}{lll}
 \hline
 \hline
 \hline
%   & \multicolumn{2}{c}{likelihood of site being in certain state} \\
state & for A sublattice & for B sublattice\\\hline
$|\!\uparrow\rangle$ or $|\!\uparrow\downarrow\rangle$ & $n_{A\uparrow} = \frac{1}{2}(n+m)$ & $n_{B\uparrow} = \frac{1}{2}(n-m)$ \\
$|\!\downarrow\rangle$ or $|\!\uparrow\downarrow\rangle$ & $n_{A\downarrow} = \frac{1}{2}(n-m)$ & $n_{B\downarrow} = \frac{1}{2}(n+m)$ \\
$|\!\uparrow\rangle$ & $n_{A\uparrow} (1- n_{A\downarrow})$ & $n_{B\uparrow} (1- n_{B\downarrow})$ \\
$|\!\downarrow\rangle$ & $n_{A\downarrow} (1- n_{A\uparrow})$ & $n_{B\downarrow} (1- n_{B\uparrow})$ \\
$|\varnothing\rangle$ & $(1-n_{A\uparrow})(1- n_{A\downarrow})$ & $(1-n_{B\uparrow})(1- n_{B\downarrow})$ \\
$|\!\uparrow\downarrow\rangle$ & $n_{A\uparrow}n_{A\downarrow}$ & $n_{B\uparrow}n_{B\downarrow}$ \\
 \hline \hline
\end{tabular}

\end{table}

In the correlated state $|\Psi \rangle$ the double occupancy $d^2$ is reduced with respect to $|\Psi_0 \rangle$ due to the presence of the onsite Coulomb repulsion.
The adjustment of $d^2$ is made by selecting a proper form of the $\hat P$ operator
(cf. Refs.~\cite{Gutzwiller1962, Gutzwiller1965}).
In the limiting case of $U \rightarrow \infty$, the double occupancies are removed in $|\Psi\rangle$, what results in $\forall_i\ \lambda_{i,d} \equiv 0$ (cf. the general form of $\hat P$, Eq.~\bref{eq:P-general}).
However, by removing the double occupancies (by changing the $\lambda_{i,d}$ parameters) we also change the average number of electrons in the system. To avoid this, the weights
$\lambda_{i,0}$, $\lambda_{i,\uparrow}$ and $\lambda_{i,\downarrow}$ need to be modified as well.

There are two intuitive ways how this can be achieved, namely
\begin{enumerate}
 \item We can ``split'' every double occupancy, separating the electrons (one $\uparrow$ and one $\downarrow$) to different,
 previously empty sites.
 Such operation would not change the global magnetization of the system
 ($m \equiv n_{\uparrow} -n_{\downarrow}$)
 but it would modify the proportion between the number of sites occupied by spin-up and down electrons.

 \item We can ``erase'' the double occupancies. However, such action would change the number of electrons in the system. Therefore, to restore the
 previous number of electrons, we can \emph{proportionally} add \emph{up} and \emph{down} electrons to previously empty sites. 
 This operation would keep the proportion of the number of single occupied states with the spin \emph{up} to those with the spin \emph{down},
 but it would modify the global magnetization of the system.
\end{enumerate}

Both of the presented schemes, lead to a different probability of sites to be in certain state, as it is displayed
in the Table~\ref{tab:Psi}. Note, that in the first scheme, the proportion of $|\!\uparrow\rangle$ states
is the same as ``$|\!\uparrow\rangle$ or $|\!\uparrow\downarrow\rangle$'' states in the Table~\ref{tab:Psi0}.
In the second scheme, after erasing the doubly occupied states, the number of the electrons has changed from $n$
to $n - n_{A\sigma} n_{A\bar\sigma}$ in the A sublattice and to $n - n_{B\sigma} n_{B\bar\sigma}$ in the B sublattice.
Therefore, to restore the previous number of electrons in the system, the probability that the state will have
a single electron $\sigma$ was renormalized by the factor $n/(n - 2 n_{A\uparrow} n_{A\downarrow}) \equiv n/(n - 2 n_{B\uparrow} n_{B\downarrow})$.

\begin{table}
\caption{Likelihood of a site being in the certain state
(correlated case $|\Psi \rangle$).
In the table, only the results for the A sublattice were shown. 
For the $B$ sublattice simply $n_{B\sigma} = n_{A\bar\sigma}$.
}\label{tab:Psi}
\centering
\begin{tabular}{lll}
 \hline
  \hline
state & scheme $1.$ (``splitting'') & scheme $2.$ (``erasing'') \\\hline
$|\!\uparrow\rangle$ & $n_{A\uparrow} = \frac{1}{2}(n+m)$ & $n_{A\uparrow}(1 - n_{A\downarrow}) \frac{n}{n - 2 n_{A\uparrow} n_{A\downarrow}}$ \\
$|\!\downarrow\rangle$ & $n_{A\downarrow} = \frac{1}{2}(n-m)$ & $n_{A\downarrow} (1 - n_{A\uparrow}) \frac{n}{n - 2 n_{A\uparrow} n_{A\downarrow}}$ \\
% $|\!\uparrow\rangle$ & $n_{A\uparrow}$ & $n_{B\uparrow} (1- n_{B\downarrow})$ \\
% $|\!\downarrow\rangle$ & $n_{A\downarrow}$ & $n_{B\downarrow} (1- n_{B\uparrow})$ \\
$|\varnothing\rangle$ & $1-n_{A\uparrow} - n_{A\downarrow}$ & $1-n_{A\uparrow} - n_{A\downarrow}$ \\
$|\!\uparrow\downarrow\rangle$ & $0$ & $0$ \\
 \hline
  \hline
\end{tabular}

\end{table}

Next, it is possible to derive the $g_t$ Gutzwiller factor for the hopping term in both schemes.
For the hopping to occur in the correlated state,
there needs to be a site occupied by a single electron with the spin $\sigma$,
while the neighboring site needs to be empty. Therefore, by 
comparing the amplitudes of the bra and the ket contributions of $\langle \Psi | \hat c_{i\sigma}^\dagger \hat c_{j\sigma} | \Psi \rangle$,
and with the help of Table~\ref{tab:Psi}, we can write that in the first scheme,
\begin{equation}
 \langle \Psi | \hat c_{i\sigma}^\dagger \hat c_{j\sigma} | \Psi \rangle \stackrel{(1)}{=} \sqrt{n_{A\sigma} n_{B\sigma}}(1-n).
\label{eq:psi-gt-sch1}
 \end{equation}
whereas in the second scheme,
\begin{equation}
 \langle \Psi | \hat c_{i\sigma}^\dagger \hat c_{j\sigma} | \Psi \rangle \stackrel{(2)}{=}
 \sqrt{\frac{n_{A\sigma}(1 - n_{A\bar\sigma}) n_{B\sigma}(1 - n_{B\bar\sigma})}{(n - 2 n_{A\uparrow} n_{A\downarrow})(n - 2 n_{B\uparrow} n_{B\downarrow})}} n(1-n).
 \label{eq:psi-gt-sch2}
\end{equation}

Analogically, we can calculate the hopping probability in the uncorrelated state.
Namely, the hopping can occur when either one site has electron with the spin $\sigma$ or it is doubly
occupied, and when either the neighboring site is empty or has one electron with the spin $\bar\sigma$ (cf. also \cite{Edegger2007}).
Using Table~\ref{tab:Psi0}, we obtain
\begin{equation}
 \langle \Psi_0 | \hat c_{i\sigma}^\dagger \hat c_{j\sigma} | \Psi_0 \rangle = \sqrt{n_{A\sigma}(1-n_{B\sigma}) n_{B\sigma}(1-n_{A\sigma})}.
 \label{eq:psi-gt-sch0}
\end{equation}
In effect, by using Eq.~\bref{app:gt_def}, we obtain either
\begin{equation}
 g_t^{(1)} = \frac{1-n}{\sqrt{(1-n_{\uparrow})(1-n_{\uparrow})}},
 \label{eq:qt-schema1}
\end{equation}
or
\begin{equation}
 g_t^{(2)} = \frac{1-n}{1 - \frac{2 n_{\uparrow} n_{\downarrow}}{n} },
 \label{eq:qt-schema2}
\end{equation}
where we denoted $n_{\sigma} \equiv n_{A\sigma} = n_{B\bar\sigma}$.
Both $g_t^{(1)}$ and $g_t^{(2)}$ are present in the literature, 
for example $g_t^{(2)}$ in Refs.~\cite{Ogawa1975-ProgTheorPhys.53.614, Zhang1988-PhysRevB.37.3759, Ogata2008-RepProgPhys, Eder1996-PhysRevB.54.R732},
whereas $g_t^{(1)}$ is identical with the zero-order renormalization factors of the DE-GWF method Refs.~\cite{Bunemann2012, Kaczmarczyk2013, Kaczmarczyk2014, Kaczmarczyk2015,Wysokinski2015-PhysRevB.92.125135}.

Note that if no AF order is present,
\begin{equation}
 g_t^{(1)} \equiv g_t^{(2)} \equiv g_t = \frac{1-n }{1 - n/2 },
\end{equation}
and there is no difference between $g_t^{(1)}$ and $g_t^{(2)}$ (cf.\ also Eq.~\bref{eq:gt-noAFnoCDW} and take $d=0$). For the case of CDW ordering (without AF), one should take in Eqs. \bref{eq:psi-gt-sch1}--\bref{eq:psi-gt-sch0}
$n_{A\sigma} = n_A$ and $n_{B\sigma} = n_{B}$, so that $n_A \neq n_B$. In such situation we have also $g_t^{(1)} \neq g_t^{(2)}$.

The above discussion, can be easily carried out for other Gutzwiller factors, that renormalize the averages from the remaining energy terms.
In effect, the results will become method dependent. 
Explicitly, it has also been checked that the two schemes lead to substantially different outputs, especially regarding the stability of the AF phase.
In the first scheme (used in the main text of this paper), AF phase is stable in the wide range of doping, from $0$ to about $\delta_{max} = 0.27$
(cf.\ Fig.~\ref{fig:CDW-V_1})
On the other hand, by using the second scheme one obtains the AF phase stability very close to the half-filling with $\delta_{max} < 0.006$
(cf. our previous paper \cite{Abram2013-PhysRevB.88.094502}).

In this paper, we have decided to use the first scheme, since it is consistent with the SGA method which in turn is equivalent to the the zeroth order DE-GWF approach.

\section*{References}
%\bibliography{bibliografia2}{}
%\bibliographystyle{iopart-num}

\end{document}